\documentclass[12pt]{article}
\pdfoutput=1
\usepackage[a4paper]{geometry}
\usepackage{jheppub, amsmath,amssymb,amsfonts,amsxtra, mathrsfs, makeidx,graphics,graphicx,amsthm,epsfig, bm,longtable,float, color,tikz,mathtools,xfrac,footnote,rotating, lscape, makecell, environ,mathtools, empheq}

\usepackage{subfig}
\usepackage{multirow}
\usepackage{adjustbox}

\usepackage{amsthm}
\usepackage{hyperref}

\usepackage{amsmath}
\usepackage{amsfonts}
\usepackage{commath}
\usepackage[english]{babel}
\usepackage{setspace}

\usepackage{epsfig,amssymb} 
\usepackage{amsmath}

\usepackage{amscd,color}
\usepackage{amsmath,graphicx}
\usepackage{verbatim,booktabs}

\usepackage{latexsym}
\usepackage{amsmath,amsfonts,amssymb,amsthm}
\usepackage{amsmath,amsthm}

\DeclareMathOperator{\Tr}{Tr}
\DeclareMathOperator{\rk}{rk}

\renewcommand{\Re}{\text{Re}}
\renewcommand{\Im}{\text{Im}}

\title{
\begin{center}
Expanding on the Cardy-like limit \\ of the SCI of 4d $\mathcal{N}=1$ ABCD SCFTs
\end{center}}

\author[a]{Antonio Amariti,}
\author[b,c]{Marco Fazzi,} 
\author[a,d]{and Alessia Segati}
\affiliation[a]{INFN, Sezione di Milano, Via Celoria 16, I-20133 Milano, Italy}
\affiliation[b]{INFN, Sezione di Milano-Bicocca, Piazza della Scienza 3, I-20126 Milano, Italy}
\affiliation[c]{Dipartimento di Fisica, Universit\`a di Milano-Bicocca, Piazza della Scienza 3, I-20126 Milano, Italy}
\affiliation[d]{Dipartimento di Fisica, Universit\`a degli Studi di Milano, Via Celoria 16, I-20133 Milano, Italy}

\emailAdd{antonio.amariti@mi.infn.it,  marco.fazzi@mib.infn.it, alessia.segati@mi.infn.it}

\abstract{
We study the Cardy-like limit of the superconformal index of generic $\mathcal{N}=1$ SCFTs with ABCD gauge algebra,  providing strong evidence for a universal formula that captures the behavior of the index at finite order in the rank and in the fugacities associated to angular momenta. The formula extends previous results valid at lowest order, and generalizes them to generic SCFTs. We corroborate the validity of our proposal by studying several examples, beyond the well-understood toric class. We compute the index also for models without a weakly-coupled gravity dual,  whose gravitational anomaly is not of order one.
}

\begin{document}

\maketitle

%
%
%
%
%
\section{Introduction and results}
\label{sec:intro}
%
%

A fascinating consequence of the holographic duality is the possibility of obtaining the black hole (BH) entropy from a state counting in the dual field theory. The first successful result in this direction was obtained in \cite{Strominger:1996sh} for asymptotically flat BHs. Many generalizations of this result have since followed. However the generalization to asymptotically AdS$_5$ BHs has been problematic and no results have been found until very recently. The reason is that the field theory quantity that is a natural candidate to encode the information about the dual BH entropy, i.e. the superconformal index (SCI) \cite{Kinney:2005ej,Romelsberger:2005eg}, does not account for the expected scaling of the degrees of freedom of the gravitational system. This mismatch follows from the large amount of cancellations between bosonic and fermionic states,  a consequence of the presence of the operator $(-1)^F$ in the index, where $F$ is the fermionic number.

This search has recently been boosted by the results of \cite{Hosseini:2017mds}, where  a field theory quantity that could reproduce the BH entropy was constructed. The authors defined an ``entropy function'' (see \cite{Sen:2007qy}) for the $\mathcal{N}=4$ SYM case, and observed that its Legendre transform  coincides with  the expected BH entropy calculated in \cite{Gutowski:2004ez,Gutowski:2004yv}.

After this result was obtained, it was soon proven that such entropy function could be obtained from the SCI, using different approaches \cite{Cabo-Bizet:2018ehj,Choi:2018hmj,Benini:2018ywd}.
These various approaches have one common denominator: they rely on an appropriate \emph{analytic continuation} of the fugacities to complex values, such as to obstruct the cancellations induced by the operator $(-1)^F$.

Many other examples and generalizations have since been worked out. It was observed that taking a Cardy-like limit of the index, i.e. considering the fugacities associated to the rotations to be very small,  one could reproduce the entropy function expected from supergravity in various classes of models with a known holographic dual description.
Furthermore it was observed that this limit is controlled by universal combinations of the coefficients of the Weyl and Euler densities, i.e. by $\Tr R$ and $\Tr R^3$, calculated in terms of an opportunely defined set of charges, that generalize the R-charges (of the matter fields) to the curved background.  These results have been extended in \cite{Honda:2019cio,ArabiArdehali:2019tdm,Cabo-Bizet:2019osg,Kim:2019yrz,Amariti:2019mgp,Cassani:2019mms,Larsen:2019oll,Lezcano:2019pae,Lanir:2019abx,Cabo-Bizet:2019eaf,Hosseini:2019iad,Goldstein:2019gpz,ArabiArdehali:2019orz,Nian:2020qsk,David:2020ems,Cabo-Bizet:2020nkr,Murthy:2020rbd,Agarwal:2020zwm,Benini:2020gjh,GonzalezLezcano:2020yeb,Copetti:2020dil,Hosseini:2020mut,Goldstein:2020yvj,Cabo-Bizet:2020ewf,Amariti:2020jyx,Lezcano:2021qbj,Benini:2021ano}.  Most relevant to our discussion, \cite{GonzalezLezcano:2020yeb} extended the Cardy-like limit to include higher orders in the expansion in the fugacities  associated to rotations (upon further identifying the two fugacities) for $\mathcal{N}=4$ $SU(N_c)$ SYM and $\mathcal{N}=1$ toric theories.  \cite{Amariti:2020jyx} did the same for the orientifold cases of $USp(2N_c)$ and $SO(N_c)$ SYM. 

In this paper we further extend the results of \cite{GonzalezLezcano:2020yeb,Amariti:2020jyx} to generic $\mathcal{N}=1$ gauge theories 
with ABCD gauge algebra. We focus on the case where the two fugacities associated to rotations are identified.  Once again we find that the index in the Cardy-like limit is controlled by the traces $\Tr R^3$ and   
$\Tr R$, weighted by two  factors that are universal in terms of the fugacity associated to the
rotation parameter.  Furthermore we find that there is a logarithmic correction related to 
the charges  of the matter fields under the center of the gauge symmetry. (We will elaborate further on this point in section \ref{sec:main}.) The main result is formula \eqref{gen}.  This result is valid both for (non-toric) theories with $\Tr R = \mathcal{O}(1)$, i.e. for models that allow a weakly-coupled gravitational dual description, and for models with $\Tr R = \mathcal{O}(N_c^2)$, with $N_c$ the rank of the gauge algebra.

We give a general proof of the formula, based on an educated guess for the solution of the saddle point equations that arise by rewriting the index as a matrix model. Moreover we study various examples to clarify the result and make the various
novel features discussed here very explicit.  Let us give a quick survey of such novelties.
\begin{itemize}
\item
First we study examples of quivers with different ranks for the various gauge groups. Namely we study the toric/non-toric Seiberg duality for the quiver associated to the $\mathbb{C}^3/\mathbb{Z}_2 \times \mathbb{Z}_2$ singularity; the dP$_4$ quiver theory; Laufer's theory \cite{Cachazo:2001gh,Aspinwall:2010mw,Collinucci:2018aho,Amariti:2019pky,Fazzi:2019gvt}. 
\item
Another interesting aspect of our analysis regards the constraints that have to be imposed on the charges. Such constraints in the toric case follow from the requirement that the superpotential has R-charge two.  In fact this requirement coincides with the anomaly freedom of the R-symmetry. In general however this is not the case; e.g.  in SQCD the superpotential vanishes and only the constraints from  anomaly cancellation are imposed.
\item
We study  models with $\Tr R$ of order $N_c^2$. Even if in these cases it is not clear what is the dual BH we are referring to, if any, we insist on using the same Cardy-like limit defined in the cases where a dual BH should exist, i.e. we use the same types of constraints on the charges. In other words, this limit may be thought of as a straightforward generalization of the Cardy-like limit of the index to models without a weakly-coupled gravity dual description.
\item
A general property of the index consists in the fact that the logarithmic corrections
are related to the degeneracy of the solutions of its saddle point equations.  We verify that 
the argument of the logarithm is obtained from the minimal value among the sums of the charges of each matter field under the centers of the factors of the product gauge group. This is in agreement with the results obtained in the toric case in \cite{GonzalezLezcano:2020yeb} and in the orientifold cases in \cite{Amariti:2020jyx}.
\end{itemize}
This paper is structured as follows. In section \ref{sec:review} we give a lighting review of the Cardy-like limit of the SCI for $\mathcal{N}=1$ gauge theories.  In section \ref{sec:main} we propose and give a formal argument supporting our main formula \eqref{gen} for the Cardy-like limit of generic $\mathcal{N}=1$ theories with ABCD gauge algebra, including finite-order corrections, generalizing preexisting results. In later sections we test and validate this formula in a series of examples: holographic $\mathcal{N}=1$ SCFTs (section \ref{sec:examplesO1}),  $\mathcal{N}=1$ SCFTs without a weakly-coupled gravity dual (section \ref{sec:examplesON}), $\mathcal{N}=2$ SCFTs (section \ref{sec:examplesN=2}). Finally we present some open questions in section \ref{sec:conc}.

%
%
\section{The Cardy-like limit of the superconformal index}
\label{sec:review}
%
%
%
%
%

In this section we expand the SCI in the Cardy-like limit in order to extract the dominant contribution and the logarithmic correction.
The SCI is defined as the trace
\begin{equation}
\mathcal{I}_\text{sc} \equiv \Tr\, (-1)^F e^{-\beta  H_{S^3 \times S^1}} p^{J_1+\frac{R}{2}} q^{J_2 + \frac{R}{2}}
\prod_{b=1}^{\rk_F} v_b^{q_b} \ ,
\end{equation}
where $J_i$ are the angular momenta on the three-sphere, $R$ is the R-charge, and $q_b$ (not to be confused with $q$) are
the conserved charges commuting with the supercharges, where the index $b$ runs over the Cartan subgroup of the flavor symmetry group $F$, $b=1,\dots, \rk_F$. The quantities $p, q$ and $v_b$ are the associated fugacities.

For a gauge theory,  the index takes the form
\begin{equation}
\label{indexgen}
\mathcal{I}_\text{sc} = \frac{(p;p)_\infty^{\rk_G} (q;q)_\infty^{\rk_G}}{|\text{Weyl}(G)|}
\oint_{T^{\rk_G}}  \prod_{i=1}^{\rk_G} \frac{dz_i}{2 \pi i z_i} 
\frac
{\prod_{I=1}^{n_\chi}
\prod_{\rho_I} 
\Gamma_e((pq)^{{R_I}/{2}} z^{\rho_I} v^{\omega _I})
}
{\prod_\alpha \Gamma_e(z^\alpha)}\ ,
\end{equation}
where $\rho_I$ ($\omega_I$) runs over the weight vectors of the representation $\mathcal{R}_I$ ($\mathcal{F}_I$) of the gauge (flavor) group of the $I$-th $\mathcal{N}=1$ matter multiplet ($n_\chi$ being their total number), and $\alpha$ runs over the simple roots of the gauge algebra with rank $\rk G$.\footnote{Hereafter we will confuse gauge algebra $\mathfrak{g}$ and group $G$, since as we will explain below the index does not capture global aspects of the latter.} The holonomies $z_i$ are defined on the unit circle,  and the index $i$ runs over the Cartan subalgebra, $i=1,\dots, \rk_G$.  The quantities $(a;b)_\infty$ are $q$-Pochhammer symbols,
$
(a;b)_\infty \equiv \prod_{k=0}^\infty (1-a b^k)
$,
and $\Gamma_e$ are elliptic Gamma functions,
\begin{equation}
\Gamma_e(z;p,q) =\Gamma_e(z) \equiv \prod_{j,k=0}^\infty 
\frac{1-p^{j+1}q^{k+1} /z}{1-p^j q^k z}\ .
\end{equation}
We then rewrite the integral formula in terms of 
modified elliptic Gamma functions $\tilde \Gamma$.
This is done by expressing the holonomies and various fugacities as
\begin{equation}
p = e^{2 \pi i \sigma},\quad
q = e^{2 \pi i \tau},\quad
v_b = e^{2 \pi i \xi_b},\quad
z_i = e^{2 \pi i u_i}
\end{equation}
with $u_i\in (0,1]$ and $u_i \sim u_i +1$.  The chemical potential of the R-symmetry is given by
\begin{equation}
v_R = \frac{1}{2}(\tau+\sigma)\ .
\end{equation}
The modified elliptic Gamma functions are then 
\begin{equation}
\tilde \Gamma(u;\tau,\sigma) 
=
\tilde \Gamma(u) 
\equiv
 \Gamma_e(e^{2 \pi i u};e^{2 \pi i \tau},e^{2 \pi i \sigma})\ ,
\end{equation}
so that the index \eqref{indexgen} becomes
\begin{equation}\label{eq:SCIgen}
\mathcal{I}_\text{sc} (\tau,\sigma,\Delta) = \frac{(p;p)_\infty^{\rk_G} (q;q)_\infty^{\rk_G}}{|\text{Weyl}(G)|}
\int  \prod_{i=1}^{\rk_G} du_i 
\frac
{ \prod_{I=1}^{n_\chi}
\prod_{\rho_I} 
\tilde \Gamma(\rho_I(\vec u) +\Delta_I )}
{\prod_\alpha \tilde \Gamma(\alpha(\vec u))}
\end{equation}
with
\begin{equation}\label{eq:chempotmatt}
\Delta_I \equiv  \omega_I(\vec \xi\,) + R_I v_R\ .
\end{equation}
There is one chemical potential $\Delta_I$ for each field in the theory, and they must 
satisfy the relations imposed by global symmetries, i.e.~each superpotential 
term is uncharged under the flavor symmetry and it has R-charge two.

Next we restrict to the case $\tau =\sigma$,\footnote{For the rest of the paper we will restrict our attention to this case only.} and expand the index in the Cardy-like limit $|\tau| \rightarrow 0$ at fixed $\arg \tau \in (0,\pi)$. In order to evaluate the index in this limit 
it is convenient to rewrite it as a matrix model by introducing the effective action $S_\text{eff}$ through
\begin{equation}\label{eq:Seff}
\mathcal{I}_\text{sc}(\tau,\Delta) \equiv \frac{1}{|\text{Weyl}(G)|} \int \prod_{i=1}^{\rk_G} du_i\,  e^{S_{\text{eff}}(\vec u;\tau,\Delta)} \ .
\end{equation}
For a model with $n_G$ gauge groups $G_a$ and a set of $n_\chi$ matter fields $\Phi$,  the effective action takes the form
\begin{align}
   S_{\text{eff}}(\vec u; \tau, \Delta) = &  \sum_{I=1 }^{n_\chi} \sum_{\rho_{I}} \log \tilde{\Gamma}\bigl(\rho_I(\vec u)+\Delta_{I} \bigr) + \sum_{a=1}^{n_G} 
    \sum_{\alpha_a} \log \theta_0 \bigl(\alpha_a(\vec u);\tau\bigr) \ + \nonumber\\
    & + \sum_{a=1}^{n_G} 2 \rk_{G_a} \log(q;q)_\infty\ .  \label{geneffact}
\end{align}
Observe that $\sum_{\rho_I} \rho_I(\vec u)$ is a formal expression that repackages the sum over the weights of the representation $\mathcal{R}_I$.  More explicitly let us consider a function $f$  and a field $\Phi_I$ in the representation  $\mathcal{R}_I$ of the gauge group: expressing the weights as $w_j(\vec u)$, where $j=1,\ldots, \dim  \mathcal{R}_I$,  we will write
\begin{equation}\label{eq:function}
f(\rho_I(\vec u)) \equiv \sum_{j=1}^{\text{dim } \mathcal{R}_I} f( w_j(\vec u))\ .
\end{equation} 
The notation $\sum_{\alpha_a} \alpha_a(\vec u)$ then refers to the (sum of the) roots of the gauge group $G_a$, i.e. the weights of the adjoint representation.  Moreover in \eqref{geneffact} we introduced the elliptic theta function
\begin{equation}
\theta_{0} (u;\tau) \equiv \prod_{k=0}^{\infty} (1-e^{2\pi i (u + k\tau)}) (1-e^{2\pi i (-u+(k+1)\tau)})\ ,
\end{equation}
which satisfies $\log \theta_0(u;\tau) = - \log \tilde{\Gamma}(u)$.\footnote{To prove this identity, one can follow the steps explained below \cite[Eq. (3.2)]{GonzalezLezcano:2020yeb}.} 

Let us now define the $\tau$-modded value of a complex $\mathbb{C} \ni u \equiv \tilde u + \tau \check u$ (with $\tilde u,\check u \in \mathbb{R}$):
\begin{equation}\label{eq:taumod}
\{u\}_\tau \equiv u -  \lfloor \Re(u)-\cot (\arg \tau)\, \Im(u) \rfloor\ ,
\end{equation}
where $ \lfloor \cdot \rfloor$ is the floor function (of a real number).  It satisfies
\begin{equation}
\{u\}_\tau = \{\tilde u\}_\tau + \tau \check{u}\ , \quad \{-u\}_\tau = \begin{cases}1- \{u\}_\tau & \tilde{u} \notin \mathbb{Z}\\-\{u\}_\tau & \tilde{u} \in \mathbb{Z}\end{cases}\ ,
\end{equation}
and for a real number $\tilde u$ it reduces to the usual modded value $\{\tilde u\} \equiv \tilde u-\lfloor \tilde u \rfloor$.  At small $|\tau|$ and fixed $\arg \tau \in (0,\pi)$ we have the following asymptotic formulae (see e.g. \cite[App. A]{GonzalezLezcano:2020yeb}):
\begin{align}
\log \,(q;q)_{\infty} = &-\frac{i \pi}{12} \left(\tau+\frac{1}{\tau}\right)
-\frac{1}{2} \log(-i \tau) + \mathcal{O}\left(e^{\frac{2 \pi \sin (\arg \tau)}{|\tau|}}\right)\ ;
 \\
\log\theta_0(u;\tau) =&\ \frac{\pi i}{\tau} \{u\}_\tau (1-\{u\}_\tau )+
\pi i  \{u\}_\tau 
-\frac{i \pi}{6 \tau}(1+3\tau +\tau^2)\ +
\nonumber \\
&+
\log\left(
\left(1-e^{-\frac{2 \pi i }{\tau}(1-\{u\}_\tau)}\right)
\left(1-e^{-\frac{2 \pi i }{\tau}(\{u\}_\tau)}\right)
\right)
+\mathcal{O}\left(
e^{\frac{2 \pi \sin (\arg \tau)}{|\tau|}}
\right)\ ;
 \\
\log\tilde \Gamma(u) 
=&\ 2 \pi i Q(\{u\}_\tau;\tau)+\mathcal{O} \left( |\tau|^{-1} e^{\frac{2 \pi \sin (\arg \tau)}{|\tau|}
\min (\{\tilde u\},1-\{\tilde u\})}  \right)\ ,
\end{align}
provided $\tilde u \nrightarrow \mathbb{Z}$. We will also need the quantity
\begin{equation}
\label{QQQ}
Q(u;\tau) \equiv -\frac{B_3(u)}{6 \tau^2} +\frac{B_2(u)}{2\tau} -\frac{5}{12}B_{1}(u) +\frac{\tau}{12}\ ,
\end{equation}
defined in terms of the Bernoulli polynomials
\begin{equation}
\label{Scardy:t1}
B_3(u)=u^3-\frac{3 }{2}u^2+\frac{1}{2}u\ , \quad B_2(u)=u^2-u+\frac{1}{6}\ , \quad B_1(u)=u-\frac{1}{2}\ .
\end{equation}
Using the above asymptotics we can expand the effective action \eqref{geneffact} in $\tau$ for small $|\tau|$, and compute its saddle point equations \emph{at leading order}:
\begin{equation}
\label{saddleeqns}
0=\frac{\partial S_\text{eff}(\vec u;\tau,\Delta)}{\partial u_{i_a}} = 
-\frac{i \pi}{\tau^2} \sum_{I=1}^{n_\chi}\sum_{\rho_I}
\frac{\partial \rho_I (\vec u)}{\partial u_{i_a}}
 B_{2}(\{ \rho_I(\vec u) + \Delta_I \}_\tau)\ ,
\end{equation}
where  $u_{i_a}$ represents the $i$-th holonomy in the $a$-th gauge group, with $i_a=1,\dots,\rk_{G_a}$ and $a=1,\dots,n_G$.
This is a set of  $ \sum_{a=1}^{n_G} \rk_{G_a}$ equations. 
We then look for solutions $\vec{u}$,  namely the saddle points of the matrix model, which contain a constant part and a linear term in $\tau$, i.e.  we make an Ansatz for the solutions of the form
\begin{equation}
\label{solns}
\vec u = \left\{ u_{i_a} = {u_*}_{i_a} + \bar u_{i_a}
\equiv  {u_*}_{i_a} + v_{i_a} \tau \ \big| \ v_{i_a} \sim \mathcal{O}(|\tau|^0) \right\}\ .
\end{equation}
We do this to capture the terms at finite order in $\tau$ in the expansion.  In fact when we plug this Ansatz back into \eqref{geneffact}, we obtain leading and subleading contributions in $\tau$, logarithmic corrections as well as finite terms. 

In the next section we will give a formula capturing all these contributions for generic $\mathcal{N}=1$ SCFTs, which only depends on the central charges $a$ and $c$, gauge algebra, and matter representations.

\section{Expansion of the index: the general formula}
\label{sec:main}
%
%
%
%
%

In the holographic case, that is for theories with a dual, the Cardy-like limit of the SCI introduced above reproduces the Legendre transform of the entropy of the dual rotating BH. Here we propose a general formula for the index in this limit at finite order both in $\rk_G$ and $\tau$, regardless of the existence of a gravity dual.  

Our main result is that the index takes the form
\begin{align}
\log \mathcal{I}_\text{sc} (\tau, \Delta) \underset{|\tau| \to 0}{=} &\ 
\frac{4 \pi  i (\eta-6 \tau +12 \eta \tau ^2+\dots) (3 c -2 a)}{27 \tau ^2}\ +
 \nonumber \\
&\ + \frac{8 \pi  i (2-5 \eta \tau+\dots)(c-a)}{6 \tau} + \log \Gamma_Z \ , \label{gen}
\end{align}
where we use the same functions $a$ and $c$ that, when evaluated on the R-charges, reproduce $\Tr R$ and $\Tr R^3$ via
$\Tr R =16(a-c) $ and $\Tr R^3 = \frac{16}{9}(5a-3c)$ \cite{Anselmi:1997am}. 
Here we evaluate these functions on a new set of {\emph{charges} $\hat \Delta_I$ (for the matter fields)  defined as 
\begin{equation}\label{eq:bracestau}
\hat \Delta_I \equiv \frac{2}{2 \tau -\eta }  \{\Delta_I\}_\tau\ ,
\end{equation}
with $\Delta_I$ defined in \eqref{eq:chempotmatt},  the $\tau$-modded value $\{ \cdot \}_\tau$ given in \eqref{eq:taumod}, and $\eta=\pm 1$.  This latter choice has been used before to study  the Cardy-like limit of the SCI \cite{Kim:2019yrz,Cabo-Bizet:2019osg} and match it against the dual BH entropy when available (see also the discussion in \cite[Sec. 5]{Hosseini:2017mds}).
 
We stress, to avoid any confusion on the interpretation of  formula \eqref{gen}, that it has to be read as follows: 
the central charges $a$ and $c$ are computed as in a generic SCFT by considering the charges of the fermions in the matter multiplets and in the vector multiplet. While for the former we use the \emph{new} charges 
$\hat \Delta_I$ defined above (instead of the R-charges), for the latter the redefinition does \emph{not} apply (and we use their R-charge).

Our aim is to  support the validity of (\ref{gen})  for 4d $\mathcal{N}=1$ SCFTs
with a generic amount of gauge groups, each with algebra of type ABCD, without specifying the ranks.  
Before proceeding  we still need to define the positive integer $\Gamma_Z$ in  (\ref{gen}). 

Let us begin by observing that the SCI cannot capture the global aspects of the gauge group.
This implies that, once the representations of the matter fields charged under the gauge group
are specified, the index of the theory with gauge group $G$ is equivalent to the index of the theory with gauge group $G/H$, where $H$ is a discrete subgroup of the center $Z(G)$.
Here we observe, in all the examples under investigation, that  the logarithmic correction to the index in the Cardy-like limit is given by $\log \Gamma_Z$, where $\Gamma_Z$ is the minimal charge of the matter fields under the center $Z(G)$ of the gauge group.
For example for $\mathcal{N}=4$ theories the matter fields are all in the adjoint representation, i.e. they have the same charge as the dimension of the center, and in this case indeed 
$\Gamma_Z = \dim Z(G)$ \cite{GonzalezLezcano:2020yeb,Amariti:2020jyx}.
On the other hand if we consider $SU(N_c)$ SQCD, the matter fields in the fundamental
representation have charge one under the gauge group, and in this case indeed we find $\Gamma_Z=1$ (see sections \ref{sub:SQCD} and \ref{sub:adjointSQCD}).
Whenever we consider models with a center symmetry given by the product $\prod_{a=1}^{n_G} \mathbb{Z}_{k_a}$, we will refer to the sum of the charges under each single $\mathbb{Z}_{k_a}$ factor as ``charge''.
With this convention in mind we can see that our definition of $\Gamma_Z$ is also consistent with what was found for toric quivers in \cite{GonzalezLezcano:2020yeb}, 
where the center symmetry is $\mathbb{Z}_{N_c}^{n_G}$
and each field $\Phi_{ij}$ is in the fundamental representation of $SU(N_c)_i$ and in the 
anti-fundamental representation of $SU(N_c)_j$,
and thus has total charge $N_c$. In this case it was indeed found that $\Gamma_Z= N_c$.

\subsection{Derivation}

We are now ready to proceed with a derivation of (\ref{gen}).
We start our analysis by focusing on  the contribution to the index of the matter fields.  For a generic field $\Phi$ we have to consider the contribution $2 \pi i Q(\{\rho_\Phi(\vec u) + \Delta_\Phi\}_\tau;\tau )$. The function $Q(u;\tau)$ was defined in \eqref{QQQ}.

\paragraph{Vanishing holonomies.}
  Let us first focus on the contribution of the field $\Phi$ at vanishing holonomies. 
 In this paper we will only be interested in a set of charges $\hat\Delta_\Phi$ for the matter fields $\Phi$ that satisfy the constraint 
 \begin{equation}
 \label{spoteq}
  \sum_{\Phi \in \mathcal{W}} \hat \Delta_\Phi = 2  \quad \Rightarrow \quad
  \sum_{\Phi \in \mathcal{W}} \{\Delta_\Phi\}_\tau = 2 \tau-\eta\ ,
 \end{equation}
 where the notation $\Phi \in \mathcal{W}$ means that we  sum over the 
 fields in \emph{each} superpotential term, i.e. \eqref{spoteq} represent a set of $n_\mathcal{W}$ (redundant) equations, where $n_\mathcal{W}$  corresponds to the number of superpotential terms. 

A field $\Phi$ then contributes to the index as
\begin{align}
2 \pi i Q(\{\Delta_\Phi\}_\tau;\tau) = &  \ 2 \pi i Q \left(  \frac{2 \tau -\eta}{2 }    \hat \Delta_\Phi +\frac{1+\eta}{2} ;  \tau \right)  \nonumber \\
=& \ 4 \pi  i \frac{ (\eta-6 \tau +12 \eta \tau ^2+\dots)}{27 \tau ^2}  (3 c (\hat \Delta_\Phi ) -2 a(\hat \Delta_\Phi )) \ +  \nonumber \\
&+ \frac{8 \pi  i}{6 \tau}  (2-5 \eta \tau+\dots) (c(\hat \Delta_\Phi )-a(\hat \Delta_\Phi )) \equiv J(\hat \Delta_\Phi)  \ ,\label{gen2}
\end{align}
consistently with (\ref{gen}).

\paragraph{Non-vanishing holonomies.}
Next, we focus on the contribution at non-vanishing holonomies.  To do that, we make the following observation. As discussed in various papers (see e.g. \cite{Kim:2019yrz,Cabo-Bizet:2019osg}), the index evaluated at zero holonomies reproduces the leading contribution to the Legendre transform of the entropy of the dual BH. Furthermore there are other saddle point solutions of  \eqref{saddleeqns} corresponding to packages of coincident holonomies placed homogeneously along the unitary circle, that reproduce the BH entropy as well.  We have checked in many concrete examples that the number of these inequivalent solutions corresponds to the integer $\Gamma_Z$ defined above.  (For instance, for an $SU(N_c)$ theory with adjoint matter fields there are $N_c$ solutions as in \cite{Cabo-Bizet:2019osg,GonzalezLezcano:2020yeb}, while in presence of fundamental matter only the solution with all the holonomies at the origin is allowed.  There are $N_c$ solutions also for toric quivers \cite{Cabo-Bizet:2019osg,GonzalezLezcano:2020yeb} because the bifundamental matter fields imply that the gauge group is  $\prod_{a=1}^{n_G} SU(N_c)_a/\mathbb{Z}_{N_c}^{\text{diag}}$.)

A crucial observation is that in each example considered here $\rho_\Phi(\vec{u}_*) \in \mathbb{Z}$, with $\vec{u}_*$ defined in (\ref{solns}). In this way we can simplify the expansion of $2 \pi i Q(\{\rho_\Phi(\vec u) + \Delta_\Phi\}_\tau;\tau )$ using that, on the saddles
\eqref{solns}, we have 
$\{ \rho_\Phi(\vec u) + \Delta_\Phi \}_\tau = \{\tau \rho_\Phi(\vec v) + \Delta_\Phi \}_\tau$.
Even if we do not have an analytic proof we expect that this result holds in general for any $\mathcal{N}=1$ SCFT. Thanks to this, we can expand the Bernoulli polynomials in terms of the holonomies as follows:
\begin{equation}
  \begin{split}
    B_3(\{v \tau+\Delta_\Phi\}_\tau) & \underset{|\tau| \to 0}{=} B_3(\{ \Delta_\Phi \}_\tau) + 3 B_2(\{ \Delta_\Phi \}_\tau) v \tau + 3 B_1( \{ \Delta_\Phi \}_\tau) v^2 \tau^2\ , \\
     B_2(\{v \tau+\Delta_\Phi\}_\tau) & \underset{|\tau| \to 0}{=} B_2(\{ \Delta_\Phi \}_\tau) + 2 B_1(\{ \Delta_\Phi \}_\tau) v \tau\ , \\
     B_1(\{v \tau+\Delta_\Phi\}_\tau) & \underset{|\tau| \to 0}{=} B_1(\{ \Delta_\Phi \}_\tau) \ .
       \end{split}
  \label{bernoulli:relations}
\end{equation}
It follows that, in the expansion of $2 \pi i Q(\{\rho_\Phi(\vec u) + \Delta_\Phi\}_\tau;\tau )$:
\begin{itemize}
\item the linear term in the holonomies vanishes for all ABCD cases;
\item the quadratic term in the holonomies corresponds to the partition function of 3d pure Chern--Simons (CS) theory at level  $-\eta T(G)$, where $T(\mathcal{R})$ is the Dynkin index of the representation $\mathcal{R}$, and $T(G)$ refers to the adjoint representation.  (See e.g. \cite[App. A]{Amariti:2020jyx} for the relevant notation about 3d CS partition functions.)

This calculation is done as follows.  We first plug the explicit form of the Bernoulli polynomials into $2 \pi i Q(\{\rho_\Phi(\vec u) + \Delta_\Phi\}_\tau;\tau )$ and expand them around the saddle points as in \eqref{bernoulli:relations}.   In this way we obtain the quadratic contributions in the variables $\vec v$, altogether amounting to
\begin{equation}
 \frac{\pi i}{2} \sum_{I=1}^{n_\chi} \rho_I^2(\vec v) (2\tau-\eta-\{ \Delta_I \}_\tau ) \ ,
 \end{equation}
which is valid for both $\eta=\pm 1$.
 The quantity  $\rho_I^2(\vec v)$  represents the sum of the squares of the weights 
 of each field $\Phi_I$ in the representation $\mathcal{R}_I$ parameterized by the holonomies $\vec v$, as explained in \eqref{eq:function}.
This can then be expressed in terms of the Dynkin index $T(\mathcal{R}_I)$.
Using this observation and the relation between $\{\Delta_{\Phi} \}_\tau$ and 
$\hat \Delta_{\Phi}$ given in \eqref{eq:bracestau} we finally find
\begin{align}
 \frac{\pi i}{2} \sum_{I=1}^{n_\chi} \rho_I^2(\vec v) (2\tau-\eta-\{ \Delta_I \}_\tau ) &=
\frac{\pi i (2\tau-\eta) }{2} \sum_{I=1}^{n_\chi}  \rho_I^2(\vec v) (1-\hat \Delta_I  )  \nonumber \\
 & =\frac{\pi i (2\tau-\eta) }{2} \! \sum_{a=1}^{n_G} \! \sum_{i_a=1}^{\rk_{G_a}}  v_{i_a}^2 \!\! \left( \! \sum_{\Phi \in G_a} \!
T(\mathcal{R}_\Phi)  (1-\hat \Delta_\Phi  )\!
\right).
\end{align}
Here the notation $\Phi \in G_a$ means that we consider the sum over \emph{all} fields 
$\Phi$ that are charged under the $a$-th gauge group $G_a$.
The leading order in $\tau$ thus reads:
\begin{equation}
 \frac{\eta \pi  }{2} \sum_{a=1}^{n_G}\sum_{i_a=1}^{\rk_{G_a}} \lambda_{i_a}^2{ \left(  \sum_{\Phi \in G_a} 
T(\mathcal{R}_\Phi)  (1-\hat \Delta_\Phi  ) \right)}\ ,
\end{equation}
where we also  defined $  \lambda_{i_a}  \equiv i  v_{i_a} $ for future convenience.

In the toric case the constraint $\sum_{\Phi \in \mathcal{W}} \hat \Delta_\Phi =2$ automatically ensures that
\begin{equation} \label{new}
T(G) +  \sum_{\Phi \in G_a}  T(\mathcal{R}_\Phi)  (\hat \Delta_\Phi -1) = 0\ ,
\end{equation}
reflecting the fact that the anomaly freedom of the R-symmetry coincides with the requirement $R(\mathcal{W})=2$.
However the relation between the constraints imposed by the superpotential and by the requirement of a non-anomalous R-symmetry does not hold in general, and we will {\bf assume} that the Cardy-like limit has to be taken by imposing  the \emph{anomaly} cancellation for the $\hat \Delta_I$ variables as well, namely condition \eqref{new} above.
\end{itemize}
Evaluating the CS integral,\footnote{By inspection we found that the integral is given by the formula
\begin{equation}
Z_{S^3}^{G_{-\eta T(G)}} = \exp \left( i \pi \frac{(|G|-\text{rk}_G)}{2}-\frac{1}{12} i \pi  (6-5 \eta )|G| \right)\ ,\nonumber
\end{equation}
where $|G|$ is the dimension of the gauge group $G\equiv \text{Lie}(\mathfrak{g})$, with $\mathfrak{g}$ of ABCD type. It would be interesting to check the validity of such a general formula for the exceptional Lie algebras as well. See \cite{Mkrtchyan:2012jh,Mkrtchyan:2013htk,Mkrtchyan:2014wia,Mkrtchyan:2020fjg} for some results in this direction.} i.e. the partition function of a pure 3d CS theory with gauge group $G$ and CS level $-\eta T(G)$, 
and summing this result to the contribution from the vector multiplets, coming from the terms in 
$\theta_0(\alpha_a(\vec u);\tau)$ and $(q;q)_\infty$ in \eqref{geneffact},
we arrive at the result
\begin{align}
\label{gen3}
& 4 \pi  i \frac{(\eta-6 \tau +12 \eta \tau ^2+\dots)}{27 \tau ^2}
 (3 c (2) -2 a(2)) +
\frac{8 \pi  i}{6 \tau} 
(2-5 \eta \tau+\dots)
(c(2)-a(2)) \nonumber \\
&=- \frac{ \pi  i  (2-5 \eta \tau+\dots)}{12 \tau} 
\end{align}
for each vector multiplet. 

In addition there is a contribution 
$\log \Gamma_Z$ coming for the degeneration of the saddle points, as discussed above.
%
%

\subsection{The examples}
In this analysis we have made an educated guess regarding the solutions of the saddle point equations  for generic matter content and gauge group, and this cannot be regarded as a rigorous proof of the formula \eqref{gen}.
For this reason and for the sake of clarity in the next section we will study some explicit examples,
supporting the result claimed in this section.
We have chosen examples that do \emph{not} belong to the vast family of toric quiver gauge theories, that have been thoroughly investigated in this context e.g. in \cite{Honda:2019cio,Amariti:2019mgp,Cabo-Bizet:2019osg,Lezcano:2019pae,Lanir:2019abx,GonzalezLezcano:2020yeb,Benini:2020gjh}.

We kick off our analysis with an exception though,  namely by studying a toric quiver gauge theory  engineered by a stack of $N_c$ D3-branes probing the $\mathbb{C}^3/\mathbb{Z}_2 \times \mathbb{Z}_2$ singularity. The reason is that in this case we can perform a Seiberg duality that lands us on a so-called \emph{non-toric} phase. We show that  matching  the Cardy-like limit of the SCI across the dual phases  requires some care in the correct identification of the matter charges under the duality map.

We then move to a fully non-toric example,  namely the quiver gauge theory corresponding to a stack of $N_c$ D3-branes probing the cone over the $\text{dP}_4$ singularity.   We choose the phase of the theory with all but one equal ranks \cite{Wijnholt:2002qz}.
We show that formula \eqref{gen} is valid in this case as well.
Another non-toric example that we tackle is Laufer's theory, introduced in \cite{Cachazo:2001gh,Aspinwall:2010mw,Collinucci:2018aho,Amariti:2019pky}.

All of these examples admit a large-$N_c$ limit with $\Tr R = \mathcal{O}(1)$, and they are conjectured to have a weakly-coupled gravity dual.  However our formula \eqref{gen} goes beyond this requirement, providing a result that should be valid also for theories without a large-$N_c$ limit dual to classical gravity. We test this conjecture by studying the case of $SU(N_c)$ SQCD and adjoint $SU(N_c)$  SQCD. In the latter theory we discuss the modification of our formalism in presence of accidental symmetries as well.  We conclude with the case of $USp(2N_c)$ SQCD and the Intriligator--Pouliot duality it enjoys.

We then move to cases with $\mathcal{N}=2$ supersymmetry. As a first example we study a family of  $\mathcal{N}=1$ Lagrangians that enhance in the infrared (IR) to the $(A_1,A_{2n-1})$ Argyres--Douglas (AD) fixed points \cite{Maruyoshi:2016tqk,Maruyoshi:2016aim,Benvenuti:2017lle,Benvenuti:2017bpg,Agarwal:2017roi}.  
We conclude our analysis with a fully Lagrangian  $\mathcal{N}=2$ SCFT with  matter fields in tensor representations: the gauge group is $SU(N_c)$ and we have a symmetric and an antisymmetric hypermultiplet. This theory is interesting both because it has a known supergravity dual description and because the matter fields force different logarithmic corrections for the even $N_c$ and odd $N_c$ case, in perfect agreement with the logic we explained below \eqref{gen}.

%
\section{\texorpdfstring{$\mathcal{N}=1$ examples with $\Tr R = \mathcal{O}(1)$}{N=1 examples with Tr R=O(1)}}
\label{sec:examplesO1}
%

\subsection{\texorpdfstring{$\mathbb{C}^3/\mathbb{Z}_2\times\mathbb{Z}_2$: a toric/non-toric duality}{C3/Z2xZ2: a toric/non-toric duality}}
\label{sub:toricnontoric}

We will start our analysis with $\mathbb{C}^3/\mathbb{Z}_2\times\mathbb{Z}_2$.
This model has a toric holographic dual description, i.e. it is obtained from  a stack of $N_c$ D3-branes probing a Calabi--Yau threefold cone over a Sasaki--Einstein five-manifolds with $U(1)^3$ isometry.
This corresponds to a toric SCFT,  described by a quiver gauge theory with four $SU(N_c)$ gauge nodes, and pairs of bifundamental and anti-bifundamental fields connecting  each pair of nodes.
There are many equivalent ways to translate the toric condition of the metric on the dual quiver.
For instance a possibility consists of \emph{planarizing} the quiver on a torus. We refer the 
reader to \cite{Kennaway:2007tq} for further details.
Even if the Cardy-like limit of  toric quiver gauge theories has been thoroughly analyzed in the literature,  we still find it useful to consider this model because by applying the rules of Seiberg duality one obtains a dual SCFT without an explicit toric description (i.e. it is not possible to represent the Seiberg-dual quiver on a two-torus). Furthermore the non-toric dual phase is instructive because the ranks of the dual gauge groups are not all equal to $N_c$.
This will be a generic feature for some other models holographically dual to non-toric manifolds that we will consider below.

\subsubsection{Toric phase}

\begin{figure}[!h]
\centering
  \includegraphics[width=0.6\textwidth]{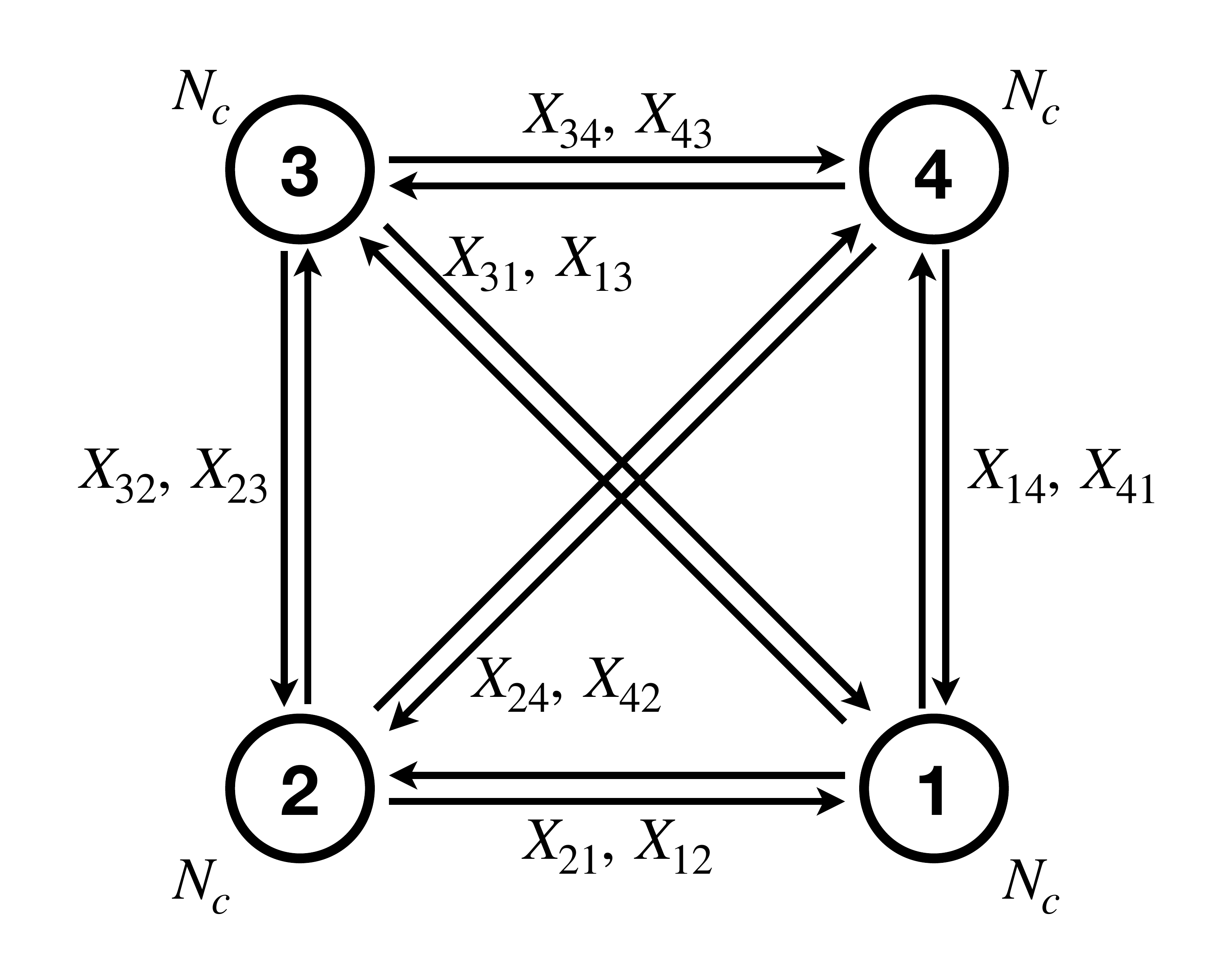}
    \caption{Quiver for the toric phase of $\mathbb{C}^3/\mathbb{Z}_2\times\mathbb{Z}_2$.}
    \label{quiver:c3z2z2}
    \end{figure}%
\noindent The toric phase is a quiver gauge theory with four gauge groups $SU(N_c)$ (see figure \ref{quiver:c3z2z2}), with all matter fields in the bifundamental representation of $SU(N_c) \times SU(N_c)$.
The superpotential reads
\begin{align}
    \mathcal{W} = &\ X_{12}X_{23}X_{31} - X_{12}X_{24}X_{41} + X_{13}X_{34}X_{41} - X_{13}X_{32}X_{21}\ + \\
    & +  X_{14}X_{42}X_{21} - X_{14}X_{43}X_{31} + X_{24}X_{43}X_{32} - X_{23}X_{34}X_{42}\ . 
  \label{superpotential:t1}
\end{align}
The SCI of this theory\footnote{The SCI for the $\mathbb{C}^3/\mathbb{Z}_k\times\mathbb{Z}_l$ orbifold theory has been computed in certain limits (albeit different from the Cardy-like one) in \cite{Bourton:2020rfo}.} is given by
\begin{equation}
    \mathcal{I}_\text{sc}(\tau,\Delta) = \frac{ (q;q)^{8N_c}_\infty}{(N_c!)^4} \int  \prod_{a=1}^{4}  \prod_{i_a=1}^{N_c} du_{i_a} \frac{\prod_{\substack{a\neq b}}^4 \prod_{i,j=1}^{N_c} \tilde{\Gamma}(u_{ij}^{ab}+\Delta_{ab})}{\prod_{a=1}^4 \prod_{ i \neq j}^{N_c}\tilde{\Gamma}(u_{ij}^a)}\ ,
  \label{SCI:t1}
\end{equation}
where $u_{i j}^{ab} \equiv u_{i_a}-u_{j_b}$ and $u_{i j}^{a} \equiv u_{i_a}-u_{j_a}$.  The effective action $S_{\text{eff}}$ \eqref{geneffact} in this case becomes:
\begin{align}
    S_{\text{eff}}(\vec u; \tau, \Delta)  =& \sum_{\substack{ a\neq b}}^4 \sum_{ i,j=1}^{N_c} \log \tilde{\Gamma}\bigl(u_{i j}^{ab}+\Delta_{ab} \bigr) + \sum_{a=1}^4 \sum_{\substack{  i \neq j}}^{N_c}  \log \theta_0 \bigl(u_{ij}^a;\tau\bigr) \ +\\
    & + 8(N_c-1)\log(q;q)_\infty\ .
  \label{Seff:t1}
\end{align}
The charges $\{ \Delta_{ab} \}_\tau$ are related by constraints that can be read off of the superpotential \eqref{superpotential:t1}, i.e. $\sum_{\Phi \in \mathcal{W}}  \{ \Delta_{\Phi} \}_\tau =2\tau-\eta$. These constraints are equivalent to those that can be derived using the anomaly cancellation for the variables $ \hat{\Delta}_{ab} \equiv \frac{2}{2\tau-\eta} \{ \Delta_{ab} \}_\tau $.
The gauge anomaly constraints on the variable $\hat \Delta_{ab}$ imply the constraints 
\begin{equation}
\label{GG}
\sum_{ \Phi \in G_a}    \{ \Delta_{\Phi} \}_\tau = 4\tau-2\eta\ , \quad a=1,\dots,4
\end{equation}
on the variables $\{\Delta_{ab}\}_\tau$. It is straightforward to prove that, in general, in the toric case the constraints \eqref{GG} are implied by the superpotential constraints.

In order to apply the saddle point approach, we need to solve the saddle point equations $\frac{\partial}{\partial u_{i_a}} S_{\text{eff}}(\vec u; \tau, \Delta) = 0$ at leading order, which for the theory at hand take the form
\begin{equation}
    -\frac{\pi i}{\tau^2} \sum_{\substack{ a\neq b}}^4 \sum_{j=1}^{N_c} \biggl( B_2 \bigl( \{ u_{ij}^{ab} + \Delta_{ab} \}_\tau \bigr) - B_2 \bigl( \{ u_{N_c j}^{ab} + \Delta_{ab} \}_\tau \bigr)  \biggr) = 0\ ,   \label{saddleseq:t1}
\end{equation}
keeping $a=1, \dots, 4$ and $i_a = 1, \dots, N_c$ fixed.

Here we consider only the solution that reproduces the BH entropy of the holographic dual description. This solution has  already been  discussed in the literature \cite{Cabo-Bizet:2019osg,GonzalezLezcano:2020yeb}. Other possible solutions, subleading in the regime of charges that we focus on, have been discussed in \cite{ArabiArdehali:2019orz,GonzalezLezcano:2020yeb}. The solution to \eqref{saddleseq:t1} is given by
\begin{equation}
  \vec{u}_a = \biggl\{ u_{j_a}^{(m)} = \frac{m}{N_c}+\bar{u}_{j_a } \equiv \frac{m}{N_c} + v_{j_a} \tau \ \Big| \ v_{j_a} \sim \mathcal{O}(|\tau|^0)\ , \ \sum_{j_a=1}^{N_c} v_{j_a}=0 \biggr\}, 
  \label{saddles:t1}
\end{equation}
with $ m=0,\dots, N_c-1$ and $a=1,\dots,4$.  We can now evaluate the effective action around its saddle points by analyzing individually the three terms in \eqref{Seff:t1}: the first is  due to the  matter fields, the second due to the  gauge fields, and the last coming from the $q$-Pochhammer symbol. 

\paragraph{Matter fields.} The contribution to the index of the matter fields is given by the Bernoulli polynomials in \eqref{Scardy:t1}, plus a term proportional to $\tau$, that is negligible in our expansion.
The Bernoulli polynomials can be simplified using the relations (\ref{bernoulli:relations}).

We notice that the terms proportional to $3B_2( \{ \Delta \}_\tau )\bar{u}$ and $2B_1( \{ \Delta \}_\tau )\bar{u}$ cancel, due to the $SU(N_c)$ constraint on the holonomies, $\sum_{i_a=1}^{N_c} \bar{u}_{i_a}=0 \mod \mathbb{Z}$. The term proportional to $3B_1( \{ \Delta \}_\tau )\bar{u}^2$ gives instead
\begin{equation}
       -\frac{\pi i}{\tau^2}  \sum_{\substack{ a\neq b}}^4 \sum_{i,j=1}^{N_c} B_1 (\{ \Delta_{ab} \}_\tau)( \bar{u}_{i j}^{ab} )^2 
   = - \frac{\pi i}{\tau^2} \eta N_c \sum_{a=1}^4 \sum_{i_a=1}^{N_c} \bar{u}_{i_a}^2 + \mathcal{O}(|\tau|),
  \label{p4:t1}
    \end{equation}
where we used both the $SU$ constraint and the constraints from the superpotential.  Finally, the parts without holonomies give the contribution 
 \begin{align}
&  -\frac{\pi i}{\tau^2} N_c^2 \sum_{\substack{ a\neq b}}^4 \biggl[\frac{1}{3} B_3 \bigl(\{ \Delta_{ab} \}_\tau \bigr)  - \tau  B_2 \bigl(\{ \Delta_{ab} \}_\tau+\frac{5}{6} \tau^2  B_1 \bigl(\{ \Delta_{ab} \}_\tau \bigr) \biggr]+\mathcal{O}(|\tau|) = \nonumber  \\
&  N_c^2  \sum_{\substack{a\neq b}}^4 J(\hat \Delta_{ab} )+ \mathcal{O}(|\tau|)\ ,
      \label{bern:t1}
 \end{align}
where $J(\hat{\Delta}_\Phi)$ was defined in \eqref{gen2}.

\paragraph{Gauge fields \& $q$-Pochhammer.} The contribution of the gauge fields  is
     \begin{equation}
  \sum_{a=1}^4 \sum_{\substack{ i \neq j}}^{N_c} \log \theta_0 \bigl(u_{ij}^a;\tau\bigr)
 =
  \sum_{a=1}^4 \sum_{\substack{  i \neq j}}^{N_c} \log \Bigl( 2 \sin \frac{\pi \bar{u}_{ij}^a}{\tau} \Bigr)-\frac{2\pi i N_c(N_c-1)}{3\tau} + \mathcal{O}(|\tau|)
       \label{gauge1111}
     \end{equation}
 while the $q$-Pochhammer symbol gives the contribution
    \begin{equation}
    8(N_c-1)\log(q;q)_\infty   =   - 4(N_c-1) \log\tau
       -\frac{2\pi i(N_c-1)}{3\tau} + 2\pi i (N_c-1) + \mathcal{O}(|\tau|)\ .
       \label{pochh2222}
     \end{equation}
     
\paragraph{Effective action \& index.}
Therefore, the Cardy-like limit of the effective action evaluated at the saddle points can be written as:
\begin{align}
     S_{\text{eff}}(\vec u; \tau, \Delta) = & -\frac{\pi i \eta}{\tau^2} N_c \sum_{a=1}^4 \sum_{i_a=1}^{N_c} \bar{u}_{i_a}^2 + \sum_{a=1}^4 \sum_{\substack{  i \neq j}}^{N_c} \log \Bigl( 2 \sin \frac{\pi \bar{u}_{ij}^a}{\tau} \Bigr) +N_c^2  \sum_{\substack{ a \neq b}}^{4} J(\hat \Delta_{ab}) + \nonumber \\
    & - \frac{2 \pi i (N_c^2-1)}{3 \tau}   + 2\pi i (N_c-1) - 4 (N_c-1) \log \tau\ ,
  \label{Ssaddle:t1}
\end{align}
with $J(\hat \Delta_\Phi)$ defined in (\ref{gen2}). The SCI is thus
\begin{align}
    \mathcal{I}_\text{sc}(\tau,\Delta)  \underset{|\tau| \to 0}{=}  &  \sum_{m=0}^{N_c-1}\frac{\mathcal{A}}{(N_c!)^4} \int \prod_{a=1}^4 \prod_{i_a=1}^{N_c} du_{i_a}\,  e^{ -\frac{\pi i \eta}{\tau^2} N_c \sum_{i_a=1}^{N_c} (\bar{u}^a_{i_a})^2 + \sum_{i \neq j}^{N_c} \log \big( 2 \sin \frac{\pi \bar{u}_{ij}^a}{\tau} \big) } \ , \label{indS:t1}
\end{align}
where the prefactor $\mathcal{A}$ is given by 
\begin{equation}
    \mathcal{A}=  e^{N_c^2 \sum_{\substack{a \neq b}}^{4} J(\hat \Delta_{ab}) - \frac{2 \pi i (N_c^2-1)}{3 \tau}   + 2\pi i (N_c-1) - 4 (N_c-1) \log \tau }\ .
  \label{A:t1}
\end{equation}
The change of variables
\begin{equation}
  \bar{u}_j =  -i \lambda_j \tau\ , \quad \sum_{j=1}^{N_c} \lambda_j=0\ ,
  \label{newvar}
\end{equation}
modifies both the measure of the integral in \eqref{indS:t1} and the contribution in $\sin \frac{\pi \bar{u}_{ij}^a}{\tau}$. In these new variables, the SCI becomes:
\begin{align}
  \mathcal{I}_\text{sc}(\tau,\Delta)  \underset{|\tau| \to 0}{=}   &  \ N_c \ e^{-2\pi i (N_c^2 - 1)} \tau^{4(N_c-1)} \cdot \nonumber \\
 &\cdot  \frac{\mathcal{A}}{(N_c!)^4} \int \prod_{a=1}^4 \prod_{i_a=1}^{N_c}d\lambda_{i_a} \ e^{ \pi i \eta N_c \sum_{j_a=1}^{N_c} \lambda_{j_a}^2 + \sum_{ j \neq k}^{N_c} \log \left(2 \sinh (\pi \lambda_{j k}^a) \right) } \ .
  \label{indSs:t1}
\end{align}
Recall that the three-sphere partition function of 3d supersymmetric $SU(N_c)_\kappa$ CS theory is given by
\begin{equation}
  Z_{{SU}(N_c)_\kappa}^\text{CS} = \frac{1}{N_c!} \int \prod_{i=1}^{N_c} d\lambda_i\  e^{ -\pi i \kappa \sum_{j=1}^{N_c} \lambda_j^2 + \sum_{ j\neq k}^{N_c}     \log \left( 2\sinh(\pi \lambda_{j k}) \right)}\ ,
  \label{ZCS}
\end{equation}
with the constraint $\sum_{j=1}^{N_c} \lambda_j=0$. 
For $\kappa=-\eta N_c$ we have
\begin{equation}
  Z_{{SU}(N_c)_{-\eta N_c}}^\text{CS}=    e^{ \frac{5}{12} i \pi  \eta  \left(N_c^2-1\right)+\frac{1}{2} i \pi  \left(N_c-1\right) N_c }
  \label{ZCSN}
\end{equation}
and thus
\begin{equation}
  \prod_{a=1}^{n_G} Z_{{SU}(N_c)_{-\eta N_c}}^\text{CS}= e^{n_G 
\left(
  \frac{5}{12} i \pi  \eta  \left(N_c^2-1\right)+\frac{1}{2} i \pi  \left(N_c-1\right) N_c \right)
}\ .
  \label{CS:nV}
\end{equation}
Using \eqref{CS:nV}, the index \eqref{indSs:t1} can be rewritten as
\begin{equation}
  \begin{split}
    & \log  \mathcal{I}_\text{sc}(\tau,\Delta) \underset{|\tau| \to 0}{=}  \log \left(N_c  \, \mathcal{A} \, e^{-2\pi i (N_c^2 - 1)} \tau^{4(N_c-1)}  e^{\pi i \frac{N_c^2-6N_c+5}{3}}  \right) \end{split}
  \label{CS:n}
\end{equation}
 The $N_c$ contribution is due to the saddle point degeneracy in \eqref{saddles:t1}, counted by $m$. 
By inspection we see that (\ref{CS:n}) coincides with the result proposed in formula  \eqref{gen}.

\subsubsection{Non-toric phase}

\begin{figure}[!h]
\centering
  \includegraphics[width=0.6\textwidth]{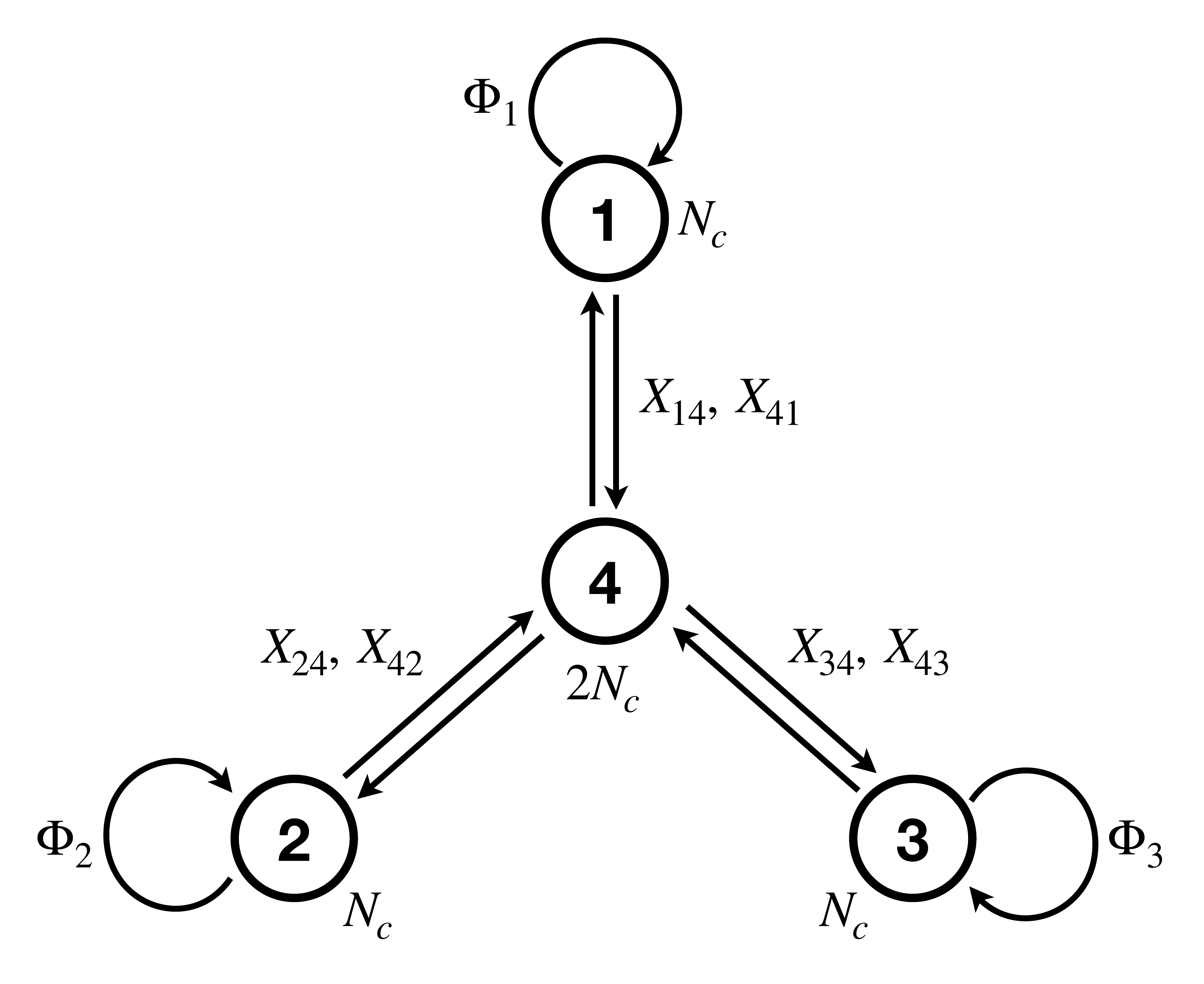}
    \caption{Quiver for the non-toric phase of $\mathbb{C}^3/\mathbb{Z}_2\times\mathbb{Z}_2$.
    Observe that the fields $\Phi_{1,2,3}$ in the figure are associated to bifundamental reducible representations. }
    \label{quiver:c3z2z2d}
    \end{figure}
   
\noindent The non-toric phase of $\mathbb{C}^3 / \mathbb{Z}_2 \times \mathbb{Z}_2$ that we are considering (figure \ref{quiver:c3z2z2d}) is obtained by performing Seiberg duality at node 4 of the original toric quiver (figure \ref{quiver:c3z2z2}).
The superpotential of the dual theory is 
\begin{equation}
  \mathcal{W}= X_{41} \Phi_1 X_{14} + X_{41} \Phi_2 X_{24} + X_{43} \Phi_3 X_{34} + X_{41}X_{14} [ X_{42}X_{24}, X_{43}X_{34}]\ ,
  \label{superpotential:nt1}
\end{equation}
where all the fields are in the bifundamental representation.

Seiberg duality implies also a mapping between the charges $\Delta_{\Phi}$ of the
electric theory (the toric phase discussed above) and the charges  $\delta_{\Phi}$ of the magnetic theory (the non-toric phase discussed here).
The explicit map is given by the following identifications:
\begin{align}
\label{dmaptnt}
    & \delta_1 = \Delta_{14}+\Delta_{41}\ , \quad  \delta_{14} = -\frac{\Delta_{41}}{2} +\frac{\Delta_{42}}{2} +\frac{\Delta_{43}}{2}\ , \quad \delta_{41} = -\frac{\Delta_{14}}{2} +\frac{\Delta_{24}}{2} +\frac{\Delta_{34}}{2}\ ,\nonumber \\
    & \delta_2 = \Delta_{24}+\Delta_{42}\ , \quad \delta_{24} = +\frac{\Delta_{41}}{2} -\frac{\Delta_{42}}{2} +\frac{\Delta_{43}}{2}\ , \quad \delta_{42} = +\frac{\Delta_{14}}{2} -\frac{\Delta_{24}}{2} +\frac{\Delta_{34}}{2}\ ,   \nonumber \\
    & \delta_3 = \Delta_{34}+\Delta_{43}\ ,\quad \delta_{34} = +\frac{\Delta_{41}}{2} +\frac{\Delta_{42}}{2} - \frac{\Delta_{43}}{2}\ , \quad \delta_{43} = + \frac{\Delta_{14}}{2} +\frac{\Delta_{24}}{2} -\frac{\Delta_{34}}{2}\ .
\end{align}
The constraints imposed by the superpotential and the anomalies on the charges $\Delta_\Phi$ automatically imply the correct constraints in the dual theory on the charges 
$\delta_{\Phi}$.
The SCI of the theory is given by:
\begin{align}
     \mathcal{I}_\text{sc}(\tau,\delta) =  & \ \frac{(q;q)^{10 N_c}_\infty}{(N_c!)^3(2N_c)!}  \int \prod_{a=1}^3 \prod_{i_a=1}^{N_c} \prod_{r=1}^{2N_c} dv_{i_a} dw_r  \prod_{a=1}^3 \left(\frac{ \prod_{i,j=1}^{N_c} \tilde{\Gamma} \bigl(v_{ij}^a + \delta_a \bigr)}{\prod_{i \neq j }^{N_c} \tilde{\Gamma} \bigl( v_{ij}^a \bigr) } \right)\cdot  \nonumber \\
      & \cdot \frac{\prod_{a=1}^3 \prod_{i_a=1}^{N_c} \prod_{r=1}^{2N_c} \tilde{\Gamma} \bigl(v_{i_a} - w_r +\delta_{a4} \bigr) \tilde{\Gamma} \bigl(w_r - v_{i_a} +\delta_{4a} \bigr)}{\prod_{r \neq s }^{2N_c}  \tilde{\Gamma} \bigl( w_{rs} \bigr) },
  \label{SCI:nt1}
\end{align}
where  $v_{ij}^a \equiv v_{i_a}-v_{j_a}$. Notice that the index $i_a = 1, \dots, N_c$ runs over the gauge groups with rank $N_c$, while the index $r=1, \dots, 2N_c$ runs over the gauge group with rank $2N_c$ (see figure \ref{quiver:c3z2z2}).
Using the relation \eqref{geneffact}, the index can be expressed in terms of the effective action, which is given by:
\begin{align}
     S_{\text{eff}}  (\vec u;\tau, \delta)  = & \sum_{a =1}^3  \sum_{i_a=1}^{N_c} \sum_{r=1}^{2N_c} \biggl( \log \tilde{\Gamma}\bigl(v_{i_a} - w_r +\delta_{a4} \bigr)  +\log \tilde{\Gamma}\bigl( w_r - v_{i_a} +\delta_{4a} \bigr) \biggr) + \nonumber \\
    &+  \sum_{a=1}^3 \sum_{i \neq j}^{N_c} \log \tilde{\Gamma}\bigl( v_{ij}^a + \delta_a \bigr) + \sum_{a=1}^3 \sum_{ i \neq j}^{N_c} \log \theta_0 \bigl(v_{ij}^a ;\tau \bigr)  +  \sum_{ r \neq s}^{2N_c} \log \theta_0 \bigl(w_{rs};\tau \bigr)  + \nonumber \\
      &+ 6(N_c-1)\log(q;q)_\infty + 2(2N_c-1)\log(q;q)_\infty\ .
  \label{Seff:nt1}
\end{align}
The constraints among the chemical potentials can be read off of the superpotential \eqref{superpotential:nt1}, with the usual relation $\sum_{\Phi \in \mathcal W} \hat \delta_\Phi = 2$
that implies $\sum_{\Phi \in \mathcal W} \{\delta_\Phi \}_\tau = 2\tau - \eta$.  Equivalently, they can be derived using the anomaly cancellation equation for the variables $\hat \delta_\Phi$. For example, for the central node of the quiver we obtain:
\begin{equation}
  2N_c + \frac{1}{2}N_c \sum_{I=1}^3(\hat \delta_{I4}+ \hat \delta_{4I}-2)=0\ ,
  \label{aneq1:nt1}
\end{equation}
and thus
\begin{equation}
\sum_{I=1}^3(\hat \delta_{I4}+ \hat \delta_{4I}) = 2
\quad
\Rightarrow
\quad
\sum_{I=1}^3(  \{ \delta_{I4} \}_\tau + \{ \delta_{4I} \}_\tau ) = 2\tau -\eta\ .
  \label{aneq3:nt1}
\end{equation}
Similar relations hold for the other gauge nodes.

Given the effective action \eqref{Seff:nt1},  the saddle point equations read:
\begin{align}
     -\frac{\pi i}{\tau^2} \biggl[ \sum_{r=1}^{2N_c} \biggl( & B_2 \bigl( \{ v_{i_a} - w_r + \delta_{a4} \}_\tau \bigr) - B_2 \bigl( \{w_r - v_{i_a} + \delta_{4a} \}_\tau \bigr)\ + \nonumber \\
      &- B_2 \bigl( \{ v_{(N_c)_a} - w_r + \delta_{a4} \}_\tau \bigr) + B_2 \bigl( \{w_r - v_{(N_c)_a} + \delta_{4a} \}_\tau \bigr) \biggr) +  \nonumber \\
      + \sum_{j_a=1}^{N_c} \biggl( & B_2 \bigl( \{ v_{ij}^a + \delta_a \}_\tau \bigr) - B_2 \bigl( \{ -v_{ij}^a + \delta_a \}_\tau \bigr)\ + \nonumber \\
        &- B_2 \bigl( \{ v_{N_c j}^a + \delta_a \}_\tau \bigr) + B_2 \bigl( \{ -v_{N_c j}^a + \delta_a \}_\tau \bigr) \biggr) \biggr] = 0\ , 
  \label{saddleseq1:nt1}
\end{align}
for $a=1,\dots,3, \   i_a=1, \dots, N_c-1$, and
\begin{align}
    -\frac{\pi i}{\tau^2} \sum_{a=1}^3 \sum_{i_a=1}^{N_c} \biggl( & -B_2 \bigl( \{ v_{i_a} - w_r + \delta_{a4} \}_\tau \bigr) + B_2 \bigl( \{ w_r - v_{i_a} + \delta_{4a} \}_\tau \bigr) + \nonumber \\
     &+ B_2 \bigl( \{ v_{i_a} - w_{2N_c} + \delta_{a4} \}_\tau \bigr) - B_2 \bigl( \{ w_{2N_c} - v_{i_a} + \delta_{4a} \}_\tau \bigr)\biggr) =0\ ,
  \label{saddleseq2:nt1}
\end{align}
for $r=1,\dots,2N_c-1$.
The leading solutions to these equations are given by
\begin{equation}
  \vec{v}_a = \biggl\{ v_{j_a}^{(m)} = \frac{m}{N_c}+\bar{v}_{j_a} \equiv \frac{m}{N_c} + \sigma_{j_a} \tau \ \Big| \ \sigma_{j_a} \sim \mathcal{O}(|\tau|^0), \ \sum_{j_a=1}^{N_c} \sigma_{j_a}=0 \biggr\}\ ,  
  \label{saddles:nt1}
\end{equation}
with $a=1,\ldots,3$ and $ m=0,\dots, N_c-1$, and
\begin{equation}
  \vec{w} = \biggl\{ w_r^{(m)} = \frac{m}{N_c}+\bar{w}_r \equiv \frac{m}{N_c} + \sigma_r \tau \ \Big| \ \sigma_r \sim \mathcal{O}(|\tau|^0), \ \sum_{r=1}^{2N_c} \sigma_r=0  \biggr\}\ ,
  \label{saddlesw:nt1}
\end{equation}
with $m=0,\dots, N_c-1$. 
As in the toric case, in order to evaluate the effective action around these saddles, we can separate the contributions coming from matter fields, gauge fields, and $q$-Pochhammer symbol.

\paragraph{Matter fields.} The relevant contribution to the index of the matter fields is given by the first three terms in \eqref{Seff:nt1}.
As in the toric case, the terms with a linear dependence from the holonomies vanish due to the $SU(N_c)$ constraint.
The terms with a quadratic dependence from the holonomies become
\begin{align}
        -\frac{\pi i}{\tau^2}  \sum_{a=1}^3 \biggl[& \sum_{i_a=1}^{N_c} \sum_{r=1}^{2N} \Bigl( B_1 (\{ \delta_{a4} \}_\tau)( \bar{v}_{i_a} -\bar{w}_r )^2 + B_1 (\{ \delta_{4a} \}_\tau)( \bar{w}_r -
        \bar{v}_{i_a} )^2 \Bigr) + \nonumber \\
         & + \sum_{i_a,j_a=1}^{N_c}  B_1 (\{ \delta_{a} \}_\tau)( \bar{v}_{i_a} -\bar{v}_{j_a} )^2 \biggr]\ ,
\end{align}
which is equal to 
\begin{equation}
-\frac{2\pi i \eta N_c}{\tau^2} \sum_{r=1}^{2N_c} \bar{w}_r^2 - \frac{\pi i \eta N_c}{\tau^2} \sum_{a=1}^3 \sum_{i_a=1}^{N_c} (\bar{v}_{i_a})^2 + \mathcal{O}(|\tau|)\ ,   \label{p4:nt1}
\end{equation}
where we have exploited the relations among the chemical potentials coming from the superpotential and the  $SU(N_c)$ condition on the holonomies.
The term without holonomies becomes:
\begin{align}
         -\frac{\pi i N_c^2}{\tau^2}  \sum_{a=1}^3  &\biggl[  \frac{1}{3}\Bigl( 2 B_3 \bigl(\{ \delta_{a4} \}_\tau \bigr) + 2 B_3 \bigl(\{ \delta_{4a} \}_\tau \bigr) + B_3 \bigl(\{ \delta_{a} \}_\tau \bigr) \Bigr)  +\nonumber \\
        & + \tau  \Bigl( 2 B_2 \bigl(\{ \delta_{a4} \}_\tau \bigr) + 2 B_2 \bigl(\{ \delta_{4a} \}_\tau \bigr) + B_2 \bigl(\{ \delta_{a} \}_\tau \bigr) \Bigr)  + \nonumber \\
        & + \frac{5}{6} \tau^2 \Bigl( 2 B_2 \bigl(\{ \delta_{a4} \}_\tau \bigr) + 2 B_2 \bigl(\{ \delta_{4a} \}_\tau \bigr) + B_2 \bigl(\{ \delta_{a} \}_\tau \bigr) \Bigr) \biggr]  + \mathcal{O}(|\tau|)
\end{align}
equal to 
\begin{equation}
N_c^2 \sum_{a=1}^3 \Bigl(2 J(\hat{\delta}_{a4}) + 2 J(\hat{\delta}_{4a}) + J(\hat{\delta}_a) \Bigr)  + \mathcal{O}(|\tau|)\ ,      \label{bern:nt1}
\end{equation}
with the definition of  $J(\hat\Delta_\Phi)$  given in \eqref{gen2}.

\paragraph{Gauge fields \& $q$-Pochhammer.} The gauge fields contribute to the index as:
 \begin{align}
     & \sum_{a=1}^3 \sum_{ i \neq j}^{N_c} \log \theta_0 \bigl(v_{ij}^a; \tau \bigr) + \sum_{r \neq s}^{2N_c} \log \theta_0 \bigl( w_{rs}; \tau \bigr) \nonumber \\
     & = \sum_{a=1}^3 \sum_{ i\neq j}^{N_c} \log \Bigl( 2 \sin \frac{\pi \bar{v}_{ij}^a}{\tau} \Bigr) + \sum_{r \neq s}^{2N_c} \log \Bigl( 2 \sin \frac{\pi \bar{w}_{rs}}{\tau} \Bigr) - \frac{\pi i(7 N_c^2 - 5 N_c )}{6\tau} + \mathcal{O}(|\tau|)\ ,
   \label{gauge:nt1}
 \end{align}
while the contribution from the $q$-Pochhammer symbol is
 \begin{equation}
  10  (N_c-1)\log(q;q)_\infty  =  -(5N_c-4) \log \tau - \frac{\pi i (5N_c-4)}{6\tau} + \frac{\pi i (5N_c-4)}{2}\ .
   \label{pochh:nt1}
 \end{equation}
 
\paragraph{Effective action \& index.}
Therefore, the Cardy-like limit of the effective action evaluated at the saddle points can be written as:
\begin{align}
     S_{\text{eff}}  (\vec u; \tau, \delta) =  & - \frac{\pi i \eta N_c}{\tau^2} \sum_{a=1}^3 \sum_{i_a=1}^{N_c} \bar{v}_{i_a}^2 - \frac{2 \pi i \eta N_c}{\tau^2}  \sum_{r=1}^{2N_c} \bar{w}_{r}^2 +   \sum_{a=1}^3 \sum_{ i \neq j}^{N_c}  \log \left( 2 \sin \frac{\pi \bar{v}_{ij}^a}{\tau} \right) + \nonumber \\
    & +  \sum_{ r \neq s}^{2N_c}  \log \Bigl( 2 \sin \frac{\pi \bar{w}_{rs}}{\tau} \Bigr) + 
N_c^2 \sum_{a=1}^3 \Bigl( 2 J(\hat{\delta}_{a4})  + 2  J(\hat{\delta}_{4a}) +  J(\hat{\delta}_{a}) \Bigr)  + \nonumber \\
    & -  \frac{7\pi i N_c^2}{6\tau}  + \frac{5}{2}\pi i N_c + \pi i \biggl( \frac{2}{3\tau} -2 \biggr) - (5N_c-4) \log\tau\ .
  \label{Ssaddle:nt1}
\end{align}
We can write the SCI of the theory expanded in the Cardy like limit  as
\begin{align}
    \mathcal{I}_\text{sc}(\tau,\delta  ) \underset{|\tau| \to 0}{=}   &\sum_{m=0}^{N_c-1}  \frac{\mathcal{A}}{(2N_c!)(N_c!)^3} \int  \prod_{a=1}^3 {\prod_{i_a=1}^{N_c} d\bar{v}_{i_a}} \ e^{ - \frac{\pi i \eta N_c}{\tau^2} \sum_{i_a=1}^{N_c} (\bar{v}_{i_a})^2 + \sum_{i \neq j}^{N_c} \log \big( 2 \sin \frac{\pi \bar{v}_{ij}^a}{\tau} \big) } \cdot \nonumber \\
    & \cdot  {\prod_{r=1}^{2N_c} d\bar{w}_r} \ e^{ - \frac{2\pi i \eta N_c}{\tau^2} \sum_{r=1}^{2N_c} \bar{w}_{r}^2 + \sum_{r \neq s}^{2N_c} \log \big( 2 \sin \frac{\pi \bar{w}_{rs}}{\tau} \big) }\ ,
  \label{indS:nt1}
\end{align}
where $\sum_{r=0}^{2N_c} \overline w_r=0$ and $\sum_{i_a=0}^{N_c} \overline v_{i_a}=0$
for $a=1,2,3$.
The prefactor $\mathcal{A}$ is given by 
\begin{equation}
    \mathcal{A}=  e^ { N_c^2  \sum_{a=1}^3 \left( 2 J(\hat{\delta}_{a4})  + 2  J(\hat{\delta}_{4a}) +  J(\hat{\delta}_{a}) \right)  + \frac{5}{2}\pi i N_c +  \pi i \left( \frac{2}{3\tau} -2 \right) -  (5N_c -4) \log\tau }\ .
  \label{A:nt1}
\end{equation}
Using the change of variables in \eqref{newvar} and the definition of the CS partition function in \eqref{ZCS}, we obtain
\begin{equation}
   \log  \mathcal{I}_\text{sc}(\tau,\delta) \!\! \! \underset{|\tau| \to 0}{=} \!\! \! \log \! \bigg( \!\! N_c  e^{- \frac{3\pi i}{2} (N_c^2 - 1)} e^{- \frac{\pi i}{2} (4 N_c^2 - 1)}  \tau^{5N_c-4} \mathcal{A} \bigg( \!\prod_{a=1}^3 \!Z_{{SU}(N_c)_{-\eta N_c}}^\text{CS} \! \! \bigg)  Z_{{SU}(2N_c)_{ -2\eta N_c}}^\text{CS} \! \bigg)  .
     \label{indSs:nt1}
\end{equation}
Performing the integral $Z_{SU(N_c)_\kappa}^\text{CS}$ as in \eqref{ZCSN} and using the duality map in (\ref{dmaptnt}) (generalizing it to the hatted charges) we  checked that 
 the index in (\ref{indSs:nt1}) coincides with the one computed in (\ref{CS:n}).
 This is a non-trivial check of the validity of our calculation in the dual non-toric phase, where the ranks of the gauge groups are not coincident.

%
%
\subsection{\texorpdfstring{Cone over $\text{dP}_4$}{Cone over dP4}}
%
%

In this section we study the Cardy-like limit of a fully non-toric quiver gauge theory,  engineered by a stack of D3-branes probing a cone over the $\text{dP}_4$ singularity.
In this case the theory has one exact R-symmetry and four non-anomalous baryonic symmetries, while there are no other flavor symmetries, reflecting the non-toricity of the model.
The quiver is reported in figure \ref{dp4q} and the superpotential is \cite{Wijnholt:2002qz}:
\begin{align}
\mathcal{W} =& \ 
a_{3,1} x_{1,4} x_{4,3}+
c_{3,1} x_{1,4} x_{4,3}+
a_{1,2} a_{3,1} x_{2,4} x_{4,3}+
a_{1,2} c_{3,1} x_{2,4} x_{4,3}+
\nonumber \\
&+
b_{1,2} c_{3,1} x_{2,4} x_{4,3}+
x_{2,4} x_{3,2} x_{4,3}+
c_{3,1} x_{1,5} x_{5,3}-
a_{3,1} b_{1,2} x_{2,5} x_{5,3}+
\nonumber \\
&-
a_{1,2} b_{3,1} x_{2,5} x_{5,3}-
b_{1,2} c_{3,1} x_{2,5} x_{5,3}+
x_{2,5} x_{3,2} x_{5,3}-
\frac{1}{2} a_{3,1} x_{1,6} x_{6,3}+
\nonumber \\
&-
\frac{1}{2} b_{3,1} x_{1,6} x_{6,3}-
\frac{1}{2} c_{3,1} x_{1,6} x_{6,3}-
\frac{1}{2} a_{1,2} a_{3,1} x_{2,6} x_{6,3}
-\frac{1}{2} a_{1,2} b_{3,1} x_{2,6} x_{6,3}+
\nonumber \\
&-
\frac{1}{2}
a_{1,2} c_{3,1} x_{2,6} x_{6,3}+
x_{2,6} x_{3,2} x_{6,3}
-
\frac{1}{2} b_{3,1} x_{1,7}x_{7,3}
-
\frac{1}{2} c_{3,1} x_{1,7}x_{7,3}+
\nonumber \\
&-
\frac{1}{2} 
a_{1,2} b_{3,1} x_{2,7} x_{7,3}
+
\frac{1}{2} 
a_{1,2} c_{3,1} x_{2,7} x_{7,3}
+
x_{2,7} x_{3,2} x_{7,3}\ .
\end{align}
\begin{figure}[!h]
\centering
  \includegraphics[width=0.6\textwidth]{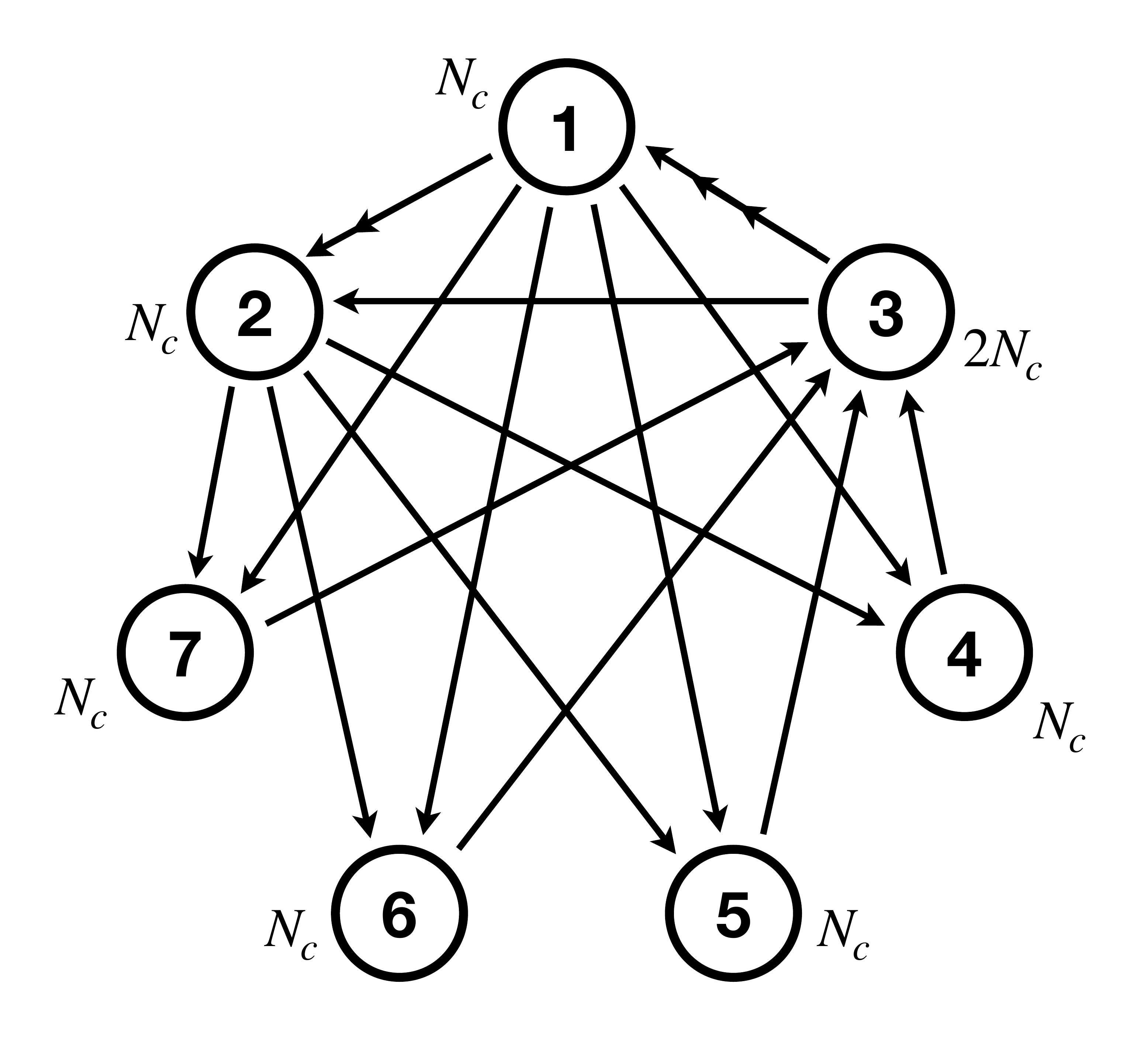}
    \caption{Quiver for cone over $\text{dP}_4$.}
    \label{dp4q}
    \end{figure}%
The table of non-anomalous global charges is
 \begin{eqnarray} 
 \label{chdP4}
\begin{array}{c|cccccccccccccccccc}
& a_{1,2} & b_{1,2} & a_{3,1} & b_{3,1} & c_{3,1} & x_{3,2} & x_{1,4} & x_{1,5} & x_{1,6} & x_{1,7} & x_{2,4} & x_{2,5} & x_{2,6} & x_{2,7} & x_{4,3} & x_{5,3} & x_{6,3} & x_{7,3} \\
\hline
U(1)_R& \frac{4}{5} & \frac{4}{5} & \frac{3}{5} & \frac{3}{5} & \frac{3}{5} & \frac{7}{5} & \frac{6}{5} & \frac{6}{5} & \frac{6}{5} & \frac{6}{5} & \frac{2}{5} & \frac{2}{5} & \frac{2}{5} & \frac{2}{5} & \frac{1}{5} & \frac{1}{5} & \frac{1}{5} & \frac{1}{5} \\
U(1)_{B_1}& 0 & 0 & 0 & 0 & 0 & 0 & 0 & 0 & 1 & -1 & 0 & 0 & 1 & -1 & 0 & 0 & -1 & 1 \\
U(1)_{B_2}& 0 & 0 & 0 & 0 & 0 & 0 & 0 & 1 & -1 & 0 & 0 & 1 & -1 & 0 & 0 & -1 & 1 & 0 \\
U(1)_{B_3}& 0 & 0 & 0 & 0 & 0 & 0 & 1 & -1 & 0 & 0 & 1 & -1 & 0 & 0 & -1 & 1 & 0 & 0 \\
U(1)_{B_4}& 4 & 4 & -2 & -2 & -2 & 2 & 1 & 1 & 1 & 1 & -3 & -3 & -3 & -3 & 1 & 1 & 1 & 1 \\
\end{array}
\nonumber
\\
\end{eqnarray}  
Let us call $q^i$ are the charges under the four non-anomalous baryonic symmetries $U(1)_{B_i}$ appearing in the table.
At the superconformal fixed point the central charges are $a= \frac{27 }{20}N_c^2-\frac{21}{16}$ and $c=\frac{27 }{20}N_c^2-\frac{7}{8}$.
    
We can now study the Cardy-like limit of the SCI, though in less detail w.r.t. the example in section \ref{sub:toricnontoric}.  Imposing the anomaly cancellation and superpotential constraints on the charges $\hat \Delta_{\Phi} $, we 
 can read the charges $\{\Delta_{\Phi} \}_\tau$ from \eqref{chdP4}:
 \begin{equation}
 \{  \Delta_\Phi \}_\tau = \frac{2\tau-\eta}{2} \bigg(  \sum_{i=1}^{4} q_\Phi^i \hat{B}_i + R_\Phi \hat{v}_R \bigg)\ ,
  \end{equation}
where the superpotential constraints imply $\hat{v}_R=1$. (This is because the model does not have any further flavor symmetry that can mix with the R-symmetry.)

We label the holonomies  as $u_{i_a}$ where $i_a=1,\dots , N_c$ for all $a=1,\dots,7$ but $3$, and $i_3 = 1,\ldots,2N_c$. These holonomies are constrained as   $\sum_{i_a=1}^{N_c} u_{i_a} =0 \mod \mathbb{Z}$  for all $a$'s but $3$, and  $\sum_{i_3=1}^{2N_c} u_{i_3} =0 \mod \mathbb{Z}$.  We then study the saddle point equations, finding the solutions
\begin{equation}
\vec{u} = \biggl\{ u_{j_a}^{(m)} = \frac{m}{N_c}+\bar{u}_{j_a} \equiv \frac{m}{N_c} + v_{j_a} \tau \ \Big| \ v_{j_a} \sim \mathcal{O}(|\tau|^0)\ , \ \sum_{j_a=1}^{N_c} v_{j_a}=0 \biggr\} \label{saddledP4alla}
\end{equation}
for all $a$'s but 3, and
\begin{equation}
\vec{u} = \biggl\{ u_{j_3}^{(m)} = \frac{m}{N_c}+\bar{u}_{j_3} \equiv \frac{m}{N_c} + v_{j_3} \tau \ \Big| \ v_{j_3} \sim \mathcal{O}(|\tau|^0)\ , \ \sum_{j_3=1}^{2 N_c} v_{j_3}=0 \biggr\}\ .  \label{saddledP4}
\end{equation}
In both cases $m=0,\dots, N_c-1$.

We then expand the effective action up to finite order in $\tau$.   In order to compute the finite-order terms we need to perform seven integrals, corresponding to the three-sphere partition function of pure CS theory $SU(2N_c)_{-2 \eta N_c} \times SU(N_c)_{- \eta N_c}^{6}$.
Evaluating these integrals we end up with the expected result (\ref{gen}).
The final result is
  \begin{equation} 
\log \mathcal{I}_\text{sc}^\text{dP$_4$}(\tau,\mathcal{B})  \underset{|\tau| \to 0}{=}  
\log N_c + \frac{ i \pi}{5}  \left(\frac{ \mathcal{B}^2 \eta  }{2 \tau ^2} +
 \frac{18  \mathcal{B}^2 +35}{ 6\tau }
+  \frac{\eta( 82   \mathcal{B}^2+175 ) }{12}
  \right)
 \end{equation} 
 where $\mathcal{B}^2\equiv N_c^2(5( \hat{B}_1^2-
 \hat{B}_2 \hat{B}_1+\hat{B}_2^2+\hat{B}_3^2-\hat{B}_2 \hat{B}_3+10 \hat{B}_4^2)-2)$.
 Observe that the $N_c$ contribution to $\mathcal{I}_\text{sc}$ is due to the 
degeneration of the solutions \eqref{saddledP4} of the saddle point equations. 

%
%
%
%
\subsection{Laufer's theory}
%
%

This model is a quiver gauge theory with product gauge group $SU(N_c) \times SU(2N_c)$ and matter content 
\begin{equation}
\begin{array}{c|cc|cc}
	& 	SU(2N_c) 		& SU(N_c)		& U(1)_R		& U(1)_B \\
	\hline
X  	&	\mathbf{Adj}			&	\mathbf{1}		& {1}/{2}	&	0	\\
Y	&	\mathbf{Adj}			&	\mathbf{1}		&{3}/{4}	&	0	\\
a	&	\overline {\mathbf{2N}_c}	&	\mathbf{N}_c		& {1}/{2}	&	1	\\
b	&	\mathbf{2N}_c			&\overline{\mathbf{N}_c}	& {1}/{2}	&	-1	\\
\end{array}
\end{equation}
The superpotential is
\begin{equation}
\mathcal{W} =X^4+ X^2 Y + X a b +(ab)^2\ .
\end{equation}
The central charges are $a=\frac{567 }{512} N_c^2-\frac{831}{2048}$ and $c=\frac{567}{512} N_c^2 -\frac{671}{2048}$. It follows that $\Tr R = \mathcal{O}(1)$. For this reason a weakly-coupled holographic dual is possible.
The metric is anyway unknown and a comparison with the gravitational result in this case is not possible at the moment. For an expanded discussion about this theory see \cite{Collinucci:2018aho,Amariti:2019pky}.

The Cardy-like limit of the SCI in this case can be studied by solving the 
saddle point equations for the holonomies $u_{i_1}$ ($i_1=1,\dots, N_c$) for the $SU(N_c)$ gauge factor,  and 
$u_{i_2}$ ($i_1=1,\dots,2N_c$) for the $SU(2N_c)$ gauge factor, with the constraints
\begin{equation}
\sum_{i_1=1}^{N_c} u_{i_1} = \sum_{i_2=1}^{2N_c} u_{i_2} = 0 \mod \mathbb{Z}\ .
\end{equation}
We find that the leading contribution in the region that we are interested in,  which corresponds to imposing the constraint
\begin{equation}
\label{oldold}
\sum_{ \Phi \in \mathcal{W}} \hat \Delta_{\Phi} = 2\ ,
\end{equation}
is given by
\begin{equation}
\vec{u} = \biggl\{ u_{j_a}^{(m)} = \frac{m}{N_c}+\bar{u}_{j_a} \equiv \frac{m}{N_c} + v_{j_a} \tau \ \Big| \ v_{j_a} \sim \mathcal{O}(|\tau|^0)\ , \ \sum_{j_a=1}^{a N_c} v_{j_a}=0 \biggr\} \label{saddleLaufer}
\end{equation}
with $a=1,2$ and $m=0,\dots,N_c-1$.
Imposing the constraints \eqref{oldold} and \eqref{new} on the charges $\hat \Delta_\Phi$
we have 
$
\{ \Delta_\Phi\}_\tau =  \frac{2 \tau - \eta}{2} (  q_\Phi \hat{B} +R_\Phi \hat{v}_R )$, or more  explicitly:
\begin{align}
& \{ \Delta_X \}_\tau = \frac{2 \tau-\eta}{4} \hat{v}_R\ ,  & \{ \Delta_a \}_\tau = \frac{2 \tau-\eta}{2} \left(  \frac{\hat{v}_R}{2} + \hat{B} \right)\ , \nonumber \\
&  \{ \Delta_Y \}_\tau  =  \frac{3 (2\tau-\eta)}{8} \hat{v}_R \ ,  &  \{ \Delta_b \}_\tau  =  \frac{2 \tau-\eta}{2} \left(  \frac{ \hat{v}_R}{2} - \hat{B} \right)\ .
\end{align}
where again the superpotential constraints imply $\hat{v}_R=1$.
Using these constraints we expand the effective action up to finite order in $\tau$.   
In order to compute the finite-order terms we need to perform two integrals, corresponding 
to the three-sphere partition function of pure CS theory $SU(2N_c)_{-2 \eta N_c} \times SU(N_c)_{ -\eta N_c}$.
Evaluating these integrals we end up with
  \begin{align}
  \log \mathcal{I}_\text{sc}^\text{Laufer}(\tau,B)  \underset{|\tau| \to 0}{=}  &\ 
\frac{1}{128} i \pi  \Bigg(\frac{4 (32 \hat{B}^2-21 ) \eta  N_c^2+13 \eta }{4 \tau ^2}+
\frac{36 (32 \hat{B}^2-21 ) N_c^2+277}{6 \tau }
\nonumber \\ 
&+ 
\frac{4}{3} N_c ((288 \hat{B}^2+11 ) \eta  N_c+144 )-\eta -128 \Bigg)
+ \log{N_c}\ , 
 \end{align} 
which matches with the expected result in \eqref{gen}.
Once again,  the $\log N_c$ contribution is recovered because of the degeneration of the solutions \eqref{saddleLaufer} to the saddle point equations.

We conclude this analysis with an observation about the solutions of the saddle point equations.
In this case the constraint $\hat{v}_R=1$ allows for another solution of the type
\begin{equation}
\vec{u} = \biggl\{ u_{j_a}^{(m_a)} = \frac{m_a}{aN_c}+\bar{u}_{j_a} \equiv \frac{m_a}{aN_c} + v_{j_a} \tau \ \Big| \ v_{j_a} \sim \mathcal{O}(|\tau|^0)\ , \ \sum_{j_a=1}^{a N_c} v_{j_a}=0 \biggr\} \label{saddleLaufer2}
\end{equation}
with $2m_1-m_2= \pm N_c$ .
This solution can be visualized on the unitary circle as follows. On the $SU(N_c)$ gauge group
the solution corresponds to placing $N_c$ holonomies at the same point $\frac{m_1}{N_c}$ with $m_1=0,\dots,N_c-1$, i.e. it is the same solution studied above.
On the other hand each value of $m_1$  fixes $m_2 = 2 m_1 \pm N_c$, where the sign is chosen such that $0<m_2<2N_c-1$.  We have checked that the index expanded around this saddle is subleading with respect to the one evaluated around \eqref{saddleLaufer}. 

%
\section{\texorpdfstring{$\mathcal{N}=1$ examples with $\Tr R = \mathcal{O}(N_c)$}{N=1 examples with Tr R=O(Nc)}}
\label{sec:examplesON}
%

%
%
%
\subsection{\texorpdfstring{$SU(N_c)$ SQCD}{SU(Nc) SQCD}}
\label{sub:SQCD}
%
%

In this section we go beyond the set of theories studied so far: we analyze SCFTs with 
$\Tr R = \mathcal{O}(N_c)$.  For this reason they are not expected to have a weakly-coupled gravity dual 
and in this sense we refer to their Cardy-like limit as a generalization of the one studied above, without  any reference to the dual rotating BH.  The simplest SCFT that we study is $SU(N_c)$ SQCD in the conformal window, with $N _f$ pairs of fundamental and antifundamental flavors denoted $Q$ and $\tilde Q$ respectively,
with $\frac{3}{2} N_f \leq N_c \leq 3N_f$.

The effective action in this case is given by the formula
\begin{align}
 S_\text{eff}^\text{SQCD}(\vec u;\tau,\Delta) =& \ 
N_f \sum_{i=1}^{N_c} \left(\log \widetilde \Gamma(u_i + \Delta_Q)+
\log \widetilde \Gamma(-u_i + \Delta_{\tilde Q})\right) + \nonumber \\
&+ \sum_{i\neq j}\log \theta_0(u_{ij};\tau)
+ 2N_c \log(q;q)_\infty\ .
\end{align}
There are $N_c-1$ saddle point equations,
\begin{equation}
\label{saddlleSQCD}
B_2(\{u_j+\Delta_Q\}_\tau)-B_2(\{u_{N_c}+\Delta_Q\}_\tau)
-
B_2(\{-u_j+\Delta_{\tilde Q}\}_\tau)+B_2(\{-u_{N_c}+\Delta_{\tilde Q}\}_\tau)= 0\ ,
\end{equation}
labeled by the index $j=1,\dots,N_c-1$.
In order to solve these equations we impose only the relation \eqref{new} on the charges 
$\hat \Delta_Q$ and $\hat \Delta_{\tilde Q}$,
obtaining
\begin{equation}
\label{constraintsSQCD}
\{ \Delta_Q  \}_\tau= \frac{2\tau-\eta}{2}  \left(1-\frac{N_c}{N_f}+\frac{\hat B}{N_c}\right)\ ,\quad
\{  \Delta_{\tilde Q}  \}_\tau=\frac{2\tau-\eta}{2}   \left(1-\frac{N_c}{N_f}-\frac{\hat B}{N_c}\right)\ .
\end{equation}
There are no further constraints to impose because the superpotential vanishes.

The leading solution of the saddle point equations (\ref{saddlleSQCD}) for the holonomies is $u_i=0$ for $i=1,\dots,N_c$. Indeed in this case we cannot package  the holonomies on the unit circle as in the cases discussed above. As expected it reflects into the fact that the fields are charged under the $\mathbb{Z}_{N_c}$  center of the gauge group with charge $1$.
We then expand the effective action evaluated around this saddle point up to finite order in $\tau$, under the constraints \eqref{constraintsSQCD} with $\eta=\pm 1$.
\begin{align}
     S_{\text{eff}}^\text{SQCD}  (\vec u; \tau, \Delta) = &  - \frac{\pi i \eta N_c}{\tau^2} \sum_{i=1}^{N_c} \bar{v}_{i}^2 
    +    \sum_{\substack{ i\neq j}}^{N_c} \log \Big( 2 \sin \frac{\pi \bar{v}_{ij}}{\tau} \Big) 
     + N_f N_c ( J(\hat{\Delta}_{Q})  + 2  J(\hat{\Delta}_{\tilde Q})  )\  + \nonumber \\
    & -\frac{i \pi  \left(\tau ^2+1\right) N_c^2}{6 \tau }+\frac{1}{2} i \pi  N_c+\frac{i \pi  \left(\tau ^2-3 \tau +1\right)}{6 \tau } - 2(N_c-1) \log\tau\ .
  \label{Ssaddle:SQCD}
\end{align}
In order to compute the finite-order terms we need to perform the matrix integral corresponding 
to the three-sphere partition function of pure 3d CS theory $SU(N_c)_{- \eta N_c}$.  Evaluating the integral  we find
\begin{align}
\label{ourSQCD}
S_\text{eff}^\text{SQCD}(\vec u;\tau,\Delta)=& \
\frac{i \pi}{6}  \Bigg(\frac{\eta  (N_f^2 (N_c^2-3 \hat{B}^2)-N_c^4)}{2 \tau ^2 N_f^2}+\frac{9 \hat{B}^2 N_f^2-2 N_c^2 N_f^2+3 N_c^4+N_f^2}{\tau  N_f^2}\ +
\nonumber \\
&+
18 \hat{B}^2 \eta -N_c \Bigg(\eta  N_c \Bigg(1-\frac{6 N_c^2}{N_f^2}\Bigg)+3\Bigg)-3
\Bigg)\ .
\end{align}
Observe the absence of corrections of the form $\log N_c$ to $\log \mathcal{I}_\text{sc}^\text{SQCD}(\tau,\Delta)$ in this case: this is because  we only have a single solution (the one with vanishing holonomies) to the saddle point equations.
Also in this case it is possible to check that the expression  \eqref{ourSQCD} matches with the
general one in \eqref{gen} in terms of the central charges $a$ and $c$ evaluated on the charges $\hat \Delta_Q$ and $\hat \Delta_{\tilde Q}$ .

We can study the Seiberg dual phase as well.  The model is $SU(N_f-N_c)$ gauge theory with $N_f$ pairs of dual fundamentals and anti-fundamentals denoted $q$ and $\tilde q$ respectively,  as well as $N_f^2$ singlets $M$, identified with the $Q \tilde Q$ mesons of the electric theory.
There is also a superpotential $\mathcal{W} = M q \tilde q$.  Imposing the constraint from the superpotential and from the anomaly cancellation on the charges $\hat \Delta_{\Phi}$ we have
\begin{align}
&\{ \Delta_q \}_\tau =\frac{2 \tau-\eta}{2} \left( \frac{N_f}{N_c} +\frac{\hat B}{N_f-N_c}\right)\ ,\nonumber \\
& \{ \Delta_{\tilde q}  \}_\tau = \frac{2 \tau-\eta}{2} \left(  \frac{N_f}{N_c} - \frac{\hat B}{N_f-N_c}\right)\ ,\\
&\{ \Delta_M  \}_\tau = (2 \tau-\eta)\left(1-\frac{N_f}{N_c}\right)\ . \nonumber
\end{align}
Using these relations, that can be regarded as the duality map, it is straightforward to match the magnetic index in the Cardy-like limit with the one obtained in the electric phase.
%
%
\subsection{\texorpdfstring{$SU(N_c)$ adjoint SQCD and accidental symmetries}{SU(Nc) adjoint SQCD and accidental symmetries}}
\label{sub:adjointSQCD}
%

In this section we study another SCFT with $\Tr R = \mathcal{O}(N_c^2)$,  namely
adjoint SQCD with a power-law superpotential for the adjoint field. The model consists of an 
$\mathcal{N}=2$ $SU(N_c)$ gauge theory with $N_f$ pairs of fundamental $Q$ and 
anti-fundamentals $\tilde Q$ chiral multiplets and an adjoint chiral multiplet $X$.
The superpotential is 
\begin{equation}
\mathcal{W} = \Tr X^{k+1}
\end{equation}
with $k \in \mathbb{N}$ and $k <N_c$.
The constraints on the charges $\hat \Delta_{\Phi}$ imply
\begin{align}
&\{ \Delta_X \}_\tau = \frac{2 \tau-\eta}{k+1}\ ,  \nonumber \\
&\{ \Delta_Q  \}_\tau = \frac{2\tau-\eta}{2}  \left(1-\frac{2N_c}{(k+1)N_f}+\frac{\hat B}{N_c}\right)\ ,  \\
&\{  \Delta_{\tilde Q}  \}_\tau =\frac{2\tau-\eta}{2}   \left(1-\frac{2N_c}{(k+1)N_f}-\frac{\hat B}{N_c}\right)\ . \nonumber
\end{align}
As in the case of SQCD the saddle point equations are solved by $u_i=0$ for $i=1,\dots,N_c$.  Performing the expansion of the effective action around this solution we obtain
\begin{align}
S_\text{eff}^\text{SQCD$_\text{adj}$} = & \frac{i \pi}{3 (k+1)^3 N_f^2}  \Bigg(\! \frac{\eta  (N_f^2 ((2 k^2+k+1) N_c^2-3 \hat{B}^2 (k+1)^2+(1-k) k)-4 N_c^4)}{2 \tau ^2}+
\nonumber \\ 
&+ \frac{N_f^2 (9 \hat{B}^2 (k+1)^2-(5 k^2+k+2) N_c^2+4 k^2-k+1)+12 N_c^4}{\tau }\ +\nonumber \\
&- \frac{\eta  (N_f^2 (36 \hat{B}^2 (k+1)^2-(k (19 k+2)+7) N_c^2+17 k^2-2 k+5)+48 N_c^4)}{2} \Bigg).  \label{ourSQCDA}
\end{align}
Again  there is no $\log N_c$ correction and it is possible to check that the expression  \eqref{ourSQCDA} matches with the
general one in \eqref{gen} in terms of the central charges $a$ and $c$ evaluated on the charges $\hat \Delta_\Phi$.

Also in this case we can study the Seiberg-dual theory, derived in \cite{Kutasov:1995ss}.
This is a $SU(kN_f-N_c)$ gauge theory  with $N_f$ pairs of fundamental $q$ and 
anti-fundamental $\tilde q$ chiral multiplets, an adjoint chiral multiplet $Y$ and $k N_f^2$ singlets
$M_j\equiv Q X^j \tilde Q$ with superpotential
\begin{equation}
\mathcal{W} = \Tr Y^{k+1} + \sum_{j=0}^{k-1} M_j q Y^{k-1-j} \tilde q\ .
\end{equation} 
In this case the constraints imposed on the charges $\hat \Delta_\Phi$ by the anomaly cancellations and by the superpotential translate into the following constraints on the $\{ \Delta_\Phi \}_\tau$ variables:
\begin{align}
&\{ \Delta_Y \}_\tau = \frac{2 \tau-\eta}{k+1}\ ,  \nonumber \\
&\{ \Delta_q  \}_\tau = \frac{2\tau-\eta}{2}  \left( \frac{2 N_c-(k-1) N_f}{(k+1) N_f}-\frac{\hat B}{k N_f-N_c}\right)\ ,\nonumber \\
&\{  \Delta_{\tilde q}  \}_\tau =\frac{2\tau-\eta}{2}   \left(\frac{2 N_c-(k-1) N_f}{(k+1) N_f}+\frac{\hat B}{k N_f-N_c}\right)\ , \\
&\{  \Delta_M  \}_\tau = (2 \tau-\eta) \Bigg( \!\! \left(1-\frac{2 N_c}{(k+1) N_f}\right)+\frac{ j}{k+1}\Bigg)\ . \nonumber 
\end{align}
Using such constraints and the fact that the saddle point equations are solved only by 
$u_i=0$ for $i=1,\dots,k N_f-N_c$, we can match the dual magnetic index in the Cardy-like limit at finite order in $\tau$.

We conclude the analysis of this model by discussing the modification of the above formulae in presence
of accidental symmetries associated to gauge-invariant operators $\mathcal{O}_i$ in the chiral ring that  violate the unitarity bound, i.e. $R_{\mathcal{O}_i} < \frac{2}{3}$ in 4d.
In this case one has to modify $a$ and $c$ accordingly \cite{Kutasov:2003iy}, by adding the contribution of a singlet
and subtracting the contribution of the operator that violates the bound. In terms of the charge $\hat \Delta_\mathcal{O}$, the variations $\bf \Delta$ underwent by the central charges are:
\begin{equation}
\label{deltaac}
{\bf \Delta} a(\hat \Delta_{\mathcal{O}}) =  a\left(\frac{2}{3}\right) -a (\hat \Delta_{\mathcal{O}})\ ,
\quad
{\bf \Delta} c(\hat \Delta_{\mathcal{O}}) =  c\left(\frac{2}{3}\right) -c (\hat \Delta_{\mathcal{O}})\ .
\end{equation}
This translates into the modification of formula \eqref{gen} by a term
\begin{align}
\label{gendelta}
& \frac{4 \pi  i (\eta-6 \tau +12 \eta \tau ^2+\dots) (3 {\bf \Delta} c(\hat \Delta_{\mathcal{O}})
  -2 {\bf \Delta} a(\hat \Delta_{\mathcal{O}}) )}{27 \tau ^2}\ +
 \nonumber \\
&+ \frac{8 \pi  i (2-5 \eta \tau+\dots)({\bf \Delta} c(\hat \Delta_{\mathcal{O}}))  - {\bf \Delta} a(\hat \Delta_{\mathcal{O}}))}{6 \tau} \ .
\end{align}
We should now modify the calculation of the SCI,  according to the discussion in \cite{Agarwal:2016pjo}, by adding the contribution
\begin{equation}\label{DeltaQ}
{\bf \Delta} Q(\mathcal{O};\tau) =  \frac{i \pi  (3 \eta  \tau ^2+4 \eta -6 \tau )}{324 \tau ^2}- Q (\{ \Delta_{\mathcal{O}} \}_\tau ;\tau )\ ,
\end{equation}
with $\hat \Delta_{\mathcal{O}}=\frac{2 }{2 \tau-\eta}  \{ \Delta_{\mathcal{O}} \}_\tau$ (and $Q(u;\tau)$ defined in \eqref{QQQ}). It is straightforward to prove that \eqref{gendelta} and \eqref{DeltaQ} coincide at leading order in $\tau$ for $\eta=\pm 1$.

These formulae can be applied to the case of adjoint SQCD, where the mesons $M_j$ hit the bound of unitarity if we vary the values of $k$, $N_c$ and $N_f$.  For example the meson $M_0$ hits the bound if
$\frac{N_f}{N_c}<\frac{6}{k+1}$. In such a case we must modify the central charges and the index as above in order to correctly reproduce its behavior in the Cardy-like limit.
On the field theory side the presence of the meson hitting the unitarity bound corresponds to 
adding a superpotential term $\Tr N(M_0+ Q \tilde Q)$ in the electric theory, where $N$ is a gauge singlet.
The first term in this superpotential is irrelevant and its coupling flows to zero in the IR while the second term becomes exactly marginal at the fixed point. 
We are then left with a free singlet of R-charge $\frac{2}{3}$ and an interacting one with R-charge $2-R_Q-R_{\tilde Q}$. This procedure, that modifies the central charge as in formula (\ref{deltaac}), clarifies also the behavior of the dual theory, where the coupling
 $M_0 q Y^{k-1} \tilde q$ becomes irrelevant and one is left with a free singlet $M_0$.
 This  analysis can be applied to match the Cardy-like limit of the electric index with the one of the dual phase if the meson $M_0$ hits the unitarity bound.
  The discussion can then be generalized to other singlets,  in this case the other mesons $M_{j>0}$.

\subsection{\texorpdfstring{$USp(2N_c)$ SQCD and Intriligator--Pouliot duality}{USp(2Nc) SQCD and Intriligator-Pouliot duality}}
\label{sub:IP}

We wish to conclude our analysis with the case of an $\mathcal{N}=1$ duality involving a real gauge group. We focus on one of the simplest cases, i.e. the generalization of Seiberg duality to $USp(2N_c)$ SQCD, originally derived in \cite{Intriligator:1995ne}.
In this case the $USp(2N_c)$  electric SQCD has $2N_f$ fundamentals $Q$ and vanishing 
superpotential and it is dual to $USp(2\tilde N_c \equiv 2(N_f-N_c-2))$ with $2N_f$ dual fundamentals $q$ and 
an antisymmetric meson $M=Q \cdot Q$ with superpotential $\mathcal{W} = M\cdot q \cdot q$, where $\cdot$ represents the symplectic product.
In this case we find that only the solution at vanishing holonomies is allowed in both the electric and magnetic theory.
The charges $\{ \Delta_\Phi\}_\tau$ are given by
\begin{equation}
\{ \Delta_Q\}_\tau = \frac{2 \tau-\eta}{2} \, \frac{N_f-N_c-1}{N_f}\ ,\quad
\{ \Delta_q\}_\tau= \frac{2 \tau-\eta}{2} - \{ \Delta_Q\}_\tau \ ,\quad
\{ \Delta_M\}_\tau =  2 \{ \Delta_Q\}_\tau \ .
\end{equation}
By performing the calculation we find that, both in the electric and magnetic
theory,  the effective action  is given by 
\begin{align}
\label{ourIPd}
 S_\text{eff}^{USp(2N_c)} (\vec u;\tau,\Delta)=&
-\frac{i \pi}{6 N_f^2}  \Bigg(\frac{\eta  (N_c+1 ) N_c ((N_c+1)^2-N_f^2 )}{\tau ^2} \ +
\nonumber \\
&
-\frac{N_c (6 (N_c+1 )^3-(4 N_c+3 ) N_f^2)}{\tau }\ +
\nonumber \\
&
+\frac{\eta  N_c (24 (N_c+1)^3-(14 N_c+9) N_f^2)}{2} \Bigg)\ ,
\end{align}
and it can be checked that $\log \mathcal{I}_\text{sc}^{USp(2N_c)}(\tau,\Delta)$ extracted from this action coincides with the general result in \eqref{gen}.

%
\section{\texorpdfstring{$\mathcal{N}=2$ examples}{N=2 examples}}
\label{sec:examplesN=2}
%

\subsection{\texorpdfstring{The $(A_1,A_{2n-1})$ Argyres--Douglas $\mathcal{N}=1$ Lagrangians}{The (A1,A{2n-1}) Argyres-Douglas N=1 Lagrangians}}
\label{sub:AD}

In this section we study a family of $\mathcal{N}=1$ Lagrangian field theories that enhance in the IR to the $\mathcal{N}=2$ $(A_1,A_{2n-1})$ AD fixed points.
 The models consist of an $\mathcal{N}=1$ $SU(n)$ gauge theory  with a fundamental $q$, an anti-fundamental $\tilde q$,
 and an adjoint $\phi$ with superpotential 
 \begin{equation}
 \mathcal{W} = \sum_{i=0}^{n-2} \alpha_i \Tr q \phi^i \tilde q+ \sum_{j=2}^{n} \beta_j \Tr \phi^j\ ,
 \end{equation}
 where  $\alpha_i$ and $\beta_j$ are gauge singlets. The table of charges is obtained by performing $a$-maximization after imposing the constraints from the anomalies and from the superpotential:
 \begin{equation}
 \begin{array}{c|ccc}
 			& U(1)_R & U(1)_T&U(1)_B\\
			\hline
q           & \frac{n+3}{3 n+3}&\frac{2 }{3 \left(n+1\right)}	&1\\
{\tilde q}	&\frac{n+3}{3 n+3}	&\frac{2 }{3 \left(n+1\right)}	&-1\\
\phi	&\frac{2}{3 \left(n+1\right)}	&\frac{2 }{3 \left(n+1\right)}	&0\\
{\alpha_i}	&\frac{4 n-2 i}{3 \left(n+1\right)}	&\frac{4 n-2 i}{3 \left(n+1\right)}	&0\\
 {\beta_j}	&2-\frac{2 j}{3 \left(n+1\right)}	&	-\frac{2 j }{3 \left(n+1\right)}&0\\
  \end{array}
\end{equation}
with $U(1)_{T,B}$ two flavor symmetries. The central charges are
\begin{equation}
a=\frac{1}{2} \left(n+1\right)+\frac{1}{2 \left(n+1\right)}-\frac{29}{24}\ ,\quad
c=\frac{1}{2} \left(n+1\right)+\frac{1}{2 \left(n+1\right)}-\frac{7}{6}\ .
\end{equation}
The Cardy-like limit is studied in terms of charges that satisfy the following constraints
\begin{align}
&\hat  \Delta_q   = \frac{n+3}{3 n+3}+\frac{2 \hat T }{3 \left(n+1\right)}+\hat B\ , \nonumber \\
&\hat   \Delta_{\tilde q}  =   \frac{n+3}{3 n+3}+\frac{2 \hat T }{3 \left(n+1\right)}-\hat B\ , \nonumber \\
&\hat   \Delta_\phi   = \frac{2}{3 \left(n+1\right)}	+\frac{2 \hat T }{3 \left(n+1\right)}\ , \label{ADconst}\\
&\hat   \Delta_{\alpha_i} =\frac{4 n-2 i}{3 \left(n+1\right)}	+\frac{2\hat T(2 n- i)}{3 \left(n+1\right)}\ ,	\nonumber \\
&\hat   \Delta_{\beta_j} = 2-\frac{2 j}{3 \left(n+1\right)}	-\frac{2 j \hat T}{3 \left(n+1\right)}\ .
\nonumber 
\end{align}
The saddle point equations are solved by $u_i=0$, consistently with what has been found in \cite{Kim:2019yrz}. Expanding the effective action around this vacuum and using the constraints
\eqref{ADconst} we obtain the following result:
\begin{align}
S_\text{eff}^\text{AD}= &-\frac{i \pi  \eta}{162 \tau ^2 (n+1)^3} && \hspace{-.25cm}  \Big[ \hat{T}^3 (n (10-n (n+1) (7 n-3))+2) + \nonumber \\
&&& -  3 \hat{T}^2 (n+1) (n ((n-6) n+2)+1) +  \nonumber \\
 &&&- 3 \hat{T} (n (n ((n-6) n+6)+6-9 \hat{B}^2 (n+1)^2)+1) +  \nonumber \\
&&& + (n+1)^2 (n (3 (9 \hat{B}^2-4) n+2)+2) \Big] + \nonumber \\
 &+ \frac{i \pi}{27 \tau  (n+1)^3}  && \hspace{-.7cm}\Big[ \hat{T}^3 (n (10-n (n+1) (7 n-3))+2) +   \nonumber \\
&   & &-  3 \hat{T}^2 (n+1) (n ((n-6) n+2)+1)+ \nonumber \\
  &&&-  3 \hat{T} n ((n-1)^2 (3 n+1)-9 \hat{B}^2 (n+1){}^2) + \nonumber \\ 
  &&&+     (n+1)^2 (n (3 (9 \hat{B}^2-4) n+5)+5) \Big] + \nonumber \\
&-\frac{i \pi  \eta}{54 (n+1)^3}  &&\hspace{-.85cm} \Big[  4 \hat{T}^3 (n (10-n (n+1) (7 n-3))+2) +    \nonumber \\ 
&&   &-  12 \hat{T}^2 (n+1) (n ((n-6) n+2)+1)+   \nonumber \\ 
  &&&+   3 \hat{T} (n (36 \hat{B}^2 (n+1)^2+(19-14 n) n^2+n+1)+1)+   \nonumber \\ 
  &&&+  (n+1)^2 (n (12 (9 \hat{B}^2-4) n+23)+23)\Big]\ ,
\end{align}
and the associated $\log \mathcal{I}_\text{sc}^\text{AD}(\tau,\Delta)$ coincides with what is expected from the general formula \eqref{gen}.
%
%
%
%
\subsection{\texorpdfstring{An $\mathcal{N}=2$ orbi-orientifold and its dual black hole entropy }{An N=2 orbi-orientifold and its dual black hole entropy}}
%
%
%
%
The last model that we study is an $\mathcal{N}=2$ SCFT with gauge group 
$SU(N_c)$, plus a symmetric and an antisymmetric hypermultiplet. This model has been studied in \cite{Bourget:2018fhe} (model A5 there) and can be engineered by $N_c$ D3-branes in the background of an O7 where a further orbifold acts on the internal spacetime \cite[Sec. 3.5]{Ennes:2000fu}.

In $\mathcal{N}=1$ language we have an adjoint chiral $X$, a symmetric chiral multiplet $S$ with its conjugate $\widetilde S$, and 
an antisymmetric chiral multiplet $A$ with its conjugate $\widetilde A$. The $\mathcal{N}=1$ superpotential is
\begin{equation}
\mathcal{W} = A X \widetilde A + S X \widetilde S\ .
\end{equation}
The table of charges is:
\begin{eqnarray}\label{symmN=2}
\begin{array}{c|cccc}
& U(1)_R&SU(2)_R&U(1)_\ell &U(1)_{\tilde \ell} \\ \hline
X & 2 & 0 & 0 & 0 \\
A& 0 & 1 & 1 & 1 \\
\widetilde A& 0 & 1 & -1 & -1 \\
S& 0 & 1 & 1 & -1 \\
\widetilde S& 0 & 1 & -1 & 1 \\
\end{array}
\end{eqnarray}
 where $U(1)_R$ and $SU(2)_R$ are the $\mathcal{N}=2$ R-symmetries (with generators $T_{\mathcal{N}=2}$ and $J_3$ respectively), and we have used the notation of \cite{Ennes:2000fu} to identify the two non-R global symmetries.
The $\mathcal{N }=1$ $U(1)_R$ R-symmetry is given by $T_{\mathcal{N}=1} = \frac{1}{3} T_{\mathcal{N }=2} + \frac{4}{3} J_3$.
We refer the reader to  \cite{Ennes:2000fu} for a more complete discussion of this model and for the holographic dual construction.


Let us make the following redefinitions: $\delta_1 \equiv \Delta_A$, $\delta_2 \equiv \Delta_S$, $\delta_3 \equiv \Delta_{\widetilde A}$, $\delta_4 \equiv \Delta_{\widetilde S}$. The SCI for this model reads:
\begin{align}
    \mathcal{I}_\text{sc}^{\mathcal{N}=2} (\tau,\Delta)= &\ \frac{(q;q)^{2(N_c-1)}_\infty \tilde{\Gamma}(\Delta_X)^{N_c-1}}{N_c!} \int \prod_{i=1}^{N_c} du_i  \frac{\prod_{a =1}^2 \prod_{i<j}^{N_c} \tilde{\Gamma}(u_{ij}^+ +\delta_a)  }{ \prod_{i \neq j}^{N_c}\tilde{\Gamma}(u_{ij}^-)} \cdot \nonumber \\
    &  \cdot \prod_{b =3}^4 \prod_{i<j}^{N_c} \tilde{\Gamma}(-u_{ij}^+ +\delta_b) \prod_{i=1}^{N_c} \tilde{\Gamma}(2u_i+\delta_2)\tilde{\Gamma}(-2u_i+\delta_4) \prod_{i\neq j}^{N_c} \tilde{\Gamma}(u_{ij}^- +\Delta_X)\ ,
  \label{SCI:CFTN=2}
\end{align}
where $u_{ij}^{\pm} \equiv u_{i}\pm u_{j}$.  Using the definition \eqref{eq:Seff}, we can write down the effective action at leading order in $|\tau|$:
\begin{align}
S_\text{eff}^{\mathcal{N}=2}(\vec u;\tau,\Delta) = & -\frac{i\pi}{3\tau^2} \Bigg( \sum_{a=1}^2 \sum_{i<j}^{N_c}  B_3(\{u_{ij}^+ + \delta_a \}_\tau)  + \sum_{a=3}^4 \sum_{i<j}^{N_c}  B_3(\{-u_{ij}^+ +\delta_a \}_\tau) \, + \nonumber \\ 
& + \sum_{i=1}^{N_c}  B_3(\{2u_i+\delta_2 \}_\tau) + B_3(\{-2u_i+\delta_4 \}_\tau) + \sum_{\substack{i\neq j}}^{N_c} B_3(\{u_{ij}^- + \Delta_X \}_\tau) \Bigg) . \label{eq:SeffN=2}
\end{align}
We now compute $\frac{\partial}{\partial u_k} S_\text{eff}^{\mathcal{N}=2} = 0$ for $k=1,\ldots,N_c-1$ upon imposing the $SU(N_c)$ constraint $\sum_{i=1}^{N_c} u_i = 0\mod \mathbb{Z}$ on the holonomies in \eqref{eq:SeffN=2}. We obtain:
\begin{align}
&\sum _{a=1}^2  \left( \sum _{i \neq k}^{N_c} B_2 (\{ u_{ik}^+ +\delta _a \}_\tau )-\sum _{i=1}^{N_c-1} B_2 ( \{ u_{i N_c}^+ +\delta _a\}_\tau)\right) + \nonumber \\
&-\sum _{a=3}^4  \left( \sum _{i\neq k}^{N_c} B_2 ( \{ -u_{ik}^+ +\delta _a \}_\tau )-\sum _{i=1}^{N_c-1} B_2 ( \{ - u_{i N_c}^+ +\delta _a \}_\tau)\right) + \nonumber \\
&+2 \Big( B_2  ( \{2 u_k+\delta_2  \}_\tau )-B_2 ( \{ -2 u_k + \delta_4  \}_\tau ) + \nonumber \\
&-B_2 (\{ 2 u_{N_c}+\delta _2  \}_\tau )+B_2 (\{ -2 u_{N_c} +\delta_4  \}_\tau) \Big) +  \nonumber \\
&+ \sum _{i=1}^{N_c} \Big( B_2 (\{ u_{ki}^- +\Delta_X \}_\tau )-B_2 ( \{  -u_{ki}^- +\Delta_X \}_\tau )+\nonumber \\ 
&-B_2( \{ u_{N_c i}^- +\Delta_X \}_\tau )+B_2(\{ -u_{N_ci}^- +\Delta_X \}_\tau ) \Big) =0 \ . \label{eq:SeffN=2eqs}
\end{align}
Given the fact that the holonomies live on a torus with modular parameter $\tau$, i.e. $u_i \sim u_i +1$, we immediately see that for even $N_c$ we can solve the above equations identically by taking all holonomies equal to $u_i = 0$ or $u_i = \frac{1}{2}$ (indeed $u_{N_c} =-\sum_{i=1}^{N_c-1} u_i = 0$ or $\frac{1}{2}(N_c-1) = \frac{1}{2} \mod 1$ respectively); for odd $N_c$ we only have the $u_i = 0$ solution (with $u_{N_c}=0$). These saddle points are again consistent with the fact that in the odd case only the gauge group $SU(N_c)$ is allowed by the matter content (i.e. it is charged under the full $\mathbb{Z}_{N_c}$ center) while in the even case we could also have $SU(N_c)/\mathbb{Z}_2$ gauge group (i.e.  the matter is uncharged under a $\mathbb{Z}_2$ subgroup of the center).  We will see that this reflects into the $\log \Gamma_Z$ correction in \eqref{gen} specialized to the current gauge group and matter content.

In order to study the Cardy-like limit of the SCI we impose the superpotential and anomaly 
constraints on the charges $\hat \Delta_\Phi$.\footnote{We also found other saddle point solutions to \eqref{eq:SeffN=2eqs}, which are however subleading in the BH region specified by these constraints on the charges $\hat \Delta_\Phi$, and for this reason are not discussed here.} These translate into the following relations on the charges $\{\Delta_\Phi\}_\tau$:
\begin{equation}
\{\Delta_\Phi\}_\tau =  \frac{2 \tau - \eta}{2} \left(R_1 \hat \Delta_1 + R_2 \hat \Delta_2 + q_\ell \hat \Delta_\ell + q_{\tilde \ell} \hat \Delta_{\tilde \ell}  \right)\ ,
\end{equation}
where $\hat{\Delta}_{1,2,\ell,\tilde{\ell}}$ are the (hatted) chemical potentials of the symmetries in table \eqref{symmN=2}, and $R_{1,2}$, $q_{\ell,\tilde{\ell}}$ the matter field charges under the latter. More  explicitly:
\begin{align}
&\{ \Delta_X\}_\tau = \frac{2\tau-\eta}{2}\hat  \Delta_1\ , \nonumber \\
&\{ \Delta_A\}_\tau = \frac{2\tau-\eta}{4} (\hat \Delta_2 +\hat \Delta_\ell+\hat\Delta_{\tilde\ell} )\ , \nonumber \\
&\{ \Delta_{\widetilde A}\}_\tau = \frac{2\tau-\eta}{4} (\hat \Delta_2 -\hat \Delta_\ell-\hat\Delta_{\tilde\ell} )\ ,  \\
&\{ \Delta_S\}_\tau= \frac{2\tau-\eta}{4} (\hat \Delta_2 +\hat \Delta_\ell-\hat\Delta_{\tilde\ell} )\ , \nonumber \\
&\{ \Delta_{\widetilde S}\}_\tau = \frac{2\tau-\eta}{4} (\hat \Delta_2 -\hat \Delta_\ell+\hat\Delta_{\tilde\ell} )\ , \nonumber 
\end{align}
where the superpotential and the anomaly cancellation impose the same constraint, namely $\hat \Delta_1+\hat \Delta_2=2$. 
We thus find:
\begin{align}
\label{oursas}
S_\text{eff}^{\mathcal{N}=2} (\vec u;\tau,\Delta)= &- \frac{i \pi  (12 \eta  \tau ^2 -6 \tau+\eta)}{32 \tau ^2} \bigg( \hat \Delta _1 (\hat\Delta _2^2-\hat \Delta _\ell^2- \hat \Delta_{\tilde \ell}^2) N_c^2-2 \hat \Delta _1 \hat \Delta _\ell \hat \Delta_{\tilde \ell} N_c \ +
\nonumber \\
&+\frac{4}{9} \hat \Delta _1 (3 \hat  \Delta _1 (\hat \Delta _2+1)-8) \bigg)
+\frac{i \pi \eta (  \hat \Delta _1-\hat \Delta _2+2)   (1-3 \tau ^2)}{72 \tau ^2}\ + \nonumber \\
&+\log \frac{3+(-1)^{N_c}}{2} \ .
\end{align}
Again we find that $\log \mathcal{I}_\text{sc}^{\mathcal{N}=2}(\tau,\Delta)$ in the Cardy-like limit is given by \eqref{gen} where the logarithmic correction is $\log 2$ in the even $N_c$ case and it vanishes in the odd $N_c$ one.  

We conclude by observing that formula \eqref{oursas}  reproduces the 
\emph{universal} result of \cite{Hosseini:2020mut} at leading order in $N_c$ for $\hat \Delta_{ \ell}= \hat \Delta_{\tilde \ell}=0$ and
up to finite order terms in $\sigma=\tau$:
\begin{equation}
S_\text{eff}^{\mathcal{N}=2} (\vec u;\tau,\Delta) = -\frac{i \pi N_c^2 (12 \eta  \tau ^2 -6 \tau+\eta) \hat \Delta _1 \hat \Delta_2^2}{32 \tau ^2} +\mathcal{O}(|\tau|)\ .
\end{equation}
We can also compute the entropy of the dual Kerr--Newman BH that is expected from the holographic duality, by considering  only the leading contribution in $N_c^2$ and distinguishing the two fugacities $\tau$ and $\sigma$ for the rotations:
\begin{equation}
S^{\mathcal{N}=2}_\text{BH} = 2 \pi  \sqrt{Q_2^2-Q_\ell^2-Q_{\tilde \ell}^2+2 Q_1 (Q_2-Q_\ell- Q_{\tilde \ell})-\frac{a}{4} (J_1+J_2)}\ ,
\end{equation}
where $a=\frac{1}{4}N_c^2$ is the central charge of the 4d theory and the other quantities are the electric charges $Q_{1,2,\ell,\tilde\ell}$ and the angular momenta $J_{1,2}$ of the dual BH. By turning off $Q_{\ell}$ and $Q_{\tilde \ell}$ this reduces to the result of \cite{Hosseini:2020mut}, as expected.

%
%
%
\section{Open questions}
\label{sec:conc}
%
%
%
%

We left open some questions that may deserve a further analysis.  First,  the validity of our formula \eqref{gen} has been claimed (and checked) only for non-exceptional gauge algebras, i.e.  the ABCD cases. In the exceptional EFG cases we did not make any claim because we are not aware of an exact evaluation of the three-sphere partition function for the associated 3d pure CS theory at level $-\eta T(G)$. (For some results in this direction see \cite{Mkrtchyan:2012jh,Mkrtchyan:2013htk,Mkrtchyan:2014wia,Mkrtchyan:2020fjg}.)
Once this integral is performed we expect our result \eqref{gen} to hold in general for all semisimple gauge algebras.

Another obvious extension of the result consists in finding the generalization of  \eqref{gen} to the case where 
$\sigma\neq \tau$,  so as to fully extend the result of \cite{Cabo-Bizet:2019osg} to finite order in both angular momenta.

Further investigations should be performed to obtain a general analysis of the saddle point equations as well.  Here we have observed through a case-by-case analysis that the number of solutions (leading to the dominant contribution to the index at $\eta=\pm 1$) is equal to the logarithm of the minimal value among the sums of the charges of each matter field under the centers of the factors of the product gauge group (or, more formally,  to the order of the character lattice of the gauge algebra modulo the action of the Weyl symmetry). It would be desirable to have a proof of such a statement. (For recent progress in this direction see \cite{Cabo-Bizet:2020nkr}.)
Furthermore we did not investigate other subleading solutions, such as the $C$-center ones discussed in \cite{GonzalezLezcano:2020yeb}. A general analysis of the subleading structure of the index is certainly an interesting subject and we are planning to come back to this point in the future.

Let us conclude this discussion with a more intricate point, that is recovering the result \eqref{gen} from the Bethe Ansatz equations (BAEs) approach, following the path laid down in \cite{Benini:2018mlo,Benini:2018ywd,Benini:2020gjh,GonzalezLezcano:2020yeb}. To make the comparison between this approach and the Cardy-like one concrete, we focused on Laufer's example.  Namely, we tried to solve the BAEs for Laufer's theory with the Ansatz $u_{i_a} = -\frac{\tau}{N_c} i_a + c_a$ ($i_a=1,\ldots, aN_c$) for the holonomies of $SU(a N_c)$ (in the product gauge group $\prod_{a=1}^2 SU(a N_c)$),\footnote{The constant $c_a$ is chosen so that $\sum_{i_a} u_{i_a}=0\mod \mathbb{Z}+\tau \mathbb{Z}$ (see \cite[Eq. (3.13)]{Benini:2018ywd}).} which closely resembles the ``basic'' solution of \cite{Hosseini:2016cyf,Hong:2018viz}, successfully used in \cite{Benini:2018ywd,Lanir:2019abx,Lezcano:2019pae,GonzalezLezcano:2020yeb} for $\mathcal{N}=4$ $SU(N_c)$ SYM and $\mathcal{N}=1$ toric quivers.  However this Ansatz does not work for Laufer's theory.  We suspect the reason behind its failure are the different ranks of the two $SU$ factors, a complication which is absent in the toric case.\footnote{Similarly, the Ansatz $u_{i_a} = -\frac{\tau}{a N_c} i_a + \tilde{c}_a$, which ``keeps track'' of the different ranks and is in spirit perhaps even closer to the basic solution, also fails to be a solution.} (Similar difficulties due to the two different ranks were encountered in \cite{Amariti:2019pky}, where a careful analysis generalizing preexisting ``rules'' was required to extract the three-sphere free energy $F_{S^3}$ of the 3d version of Laufer's theory.)

A possible way to circumvent this difficulty and find the right Ansatz in cases with $SU$ groups of different ranks (if a general one exists at all), is to start from the toric/non-toric duality of section \ref{sub:toricnontoric}. Indeed we know that in the toric phase the basic Ansatz must be a solution to the BAEs (due to the results of \cite{Lanir:2019abx,Lezcano:2019pae}). On the other hand, the solution to the BAEs of the Seiberg-dual non-toric phase (which features different ranks -- see figure \ref{quiver:c3z2z2d}) can be explicitly identified by using the duality map of \cite[Sec. 6.4]{Closset:2017bse}, which proved the BAEs ``respect'' Seiberg duality for $SU(N_c)$ gauge theories (i.e. dual BA operators are equal when evaluated on dual solutions).

This example shows that a direct comparison between our formula \eqref{gen} and the analog extracted from the BAEs approach is therefore not an easy feat. A complete analysis of this and related issues is left for future work.

\section*{Acknowledgments}
We wish to thank A.~Nedelin and A.~Zaffaroni for useful discussions. This work has been supported in part by
the Italian Ministero dell’Istruzione, Universit\`a e Ricerca (MIUR), in part by Istituto
Nazionale di Fisica Nucleare (INFN) through the ``Gauge Theories, Strings, Supergravity'' (GSS) research project and in part by MIUR-PRIN contract 2017CC72MK003. The work of M.F. is supported in part by the European Union's Horizon 2020
research and innovation programme under the Marie Skłodowska-Curie grant agreement No.~754496~-~FELLINI.

\bibliographystyle{JHEP}
\bibliography{ref}

\end{document}